\pdfoutput=1
\RequirePackage{ifpdf}
\ifpdf 
\documentclass[pdftex]{sigma}
\else
\documentclass{sigma}
\fi

\numberwithin{equation}{section}

\newtheorem{Theorem}{Theorem}[section]
\newtheorem*{Theorem*}{Theorem}

\newtheorem{Lemma}[Theorem]{Lemma}

\newtheorem{rhp}[Theorem]{Riemann--Hilbert Problem}
 { \theoremstyle{definition}

\newtheorem{Remark}[Theorem]{Remark} }

\begin{document}
\allowdisplaybreaks

\renewcommand{\thefootnote}{}

\newcommand{\arXivNumber}{2308.16051}

\renewcommand{\PaperNumber}{008}

\FirstPageHeading

\ShortArticleName{Algebraic Solutions of the Painlev\'e-III $({\rm D}_7)$ Equation}

\ArticleName{Differential Equations for Approximate Solutions\\ of Painlev\'e Equations: Application to the Algebraic\\ Solutions of the Painlev\'e-III $\boldsymbol{({\rm D}_7)}$ Equation\footnote{This paper is a~contribution to the Special Issue on Evolution Equations, Exactly Solvable Models and Random Matrices in honor of Alexander Its' 70th birthday. The~full collection is available at \href{https://www.emis.de/journals/SIGMA/Its.html}{https://www.emis.de/journals/SIGMA/Its.html}}}

\Author{Robert J. BUCKINGHAM~$^{\rm a}$ and Peter D.~MILLER~$^{\rm b}$}

\AuthorNameForHeading{R.J.~Buckingham and P.D.~Miller}

\Address{$^{\rm a)}$~Department of Mathematical Sciences, University of Cincinnati,\\
\hphantom{$^{\rm a)}$}~P.O.~Box 210025, Cincinnati, OH 45221, USA}
\EmailD{\href{mailto:buckinrt@uc.edu}{buckinrt@uc.edu}}

\Address{$^{\rm b)}$~Department of Mathematics, University of Michigan,\\
\hphantom{$^{\rm b)}$}~East Hall, 530 Church St., Ann Arbor, MI 48109, USA}
\EmailD{\href{mailto:millerpd@umich.edu}{millerpd@umich.edu}}

\ArticleDates{Received August 31, 2023, in final form January 05, 2024; Published online January 20, 2024}

\Abstract{It is well known that the Painlev\'e equations can formally degenerate to autonomous differential equations with elliptic function solutions in suitable scaling limits. A~way to make this degeneration rigorous is to apply Deift--Zhou steepest-descent techniques to a Riemann--Hilbert representation of a~family of solutions. This method leads to an explicit approximation formula in terms of theta functions and related algebro-geometric ingredients that is difficult to directly link to the expected limiting differential equation. However, the approximation arises from an outer parametrix that satisfies relatively simple conditions. By applying a method that we learned from Alexander Its, it is possible to use these simple conditions to directly obtain the limiting differential equation, bypassing the details of the algebro-geometric solution of the outer parametrix problem. In this paper, we illustrate the use of this method to relate an approximation of the algebraic solutions of the Painlev\'e-III (D$_7$) equation valid in the part of the complex plane where the poles and zeros of the solutions asymptotically reside to a~form of the Weierstra\ss\ equation.}


\Classification{34E05; 34M55; 37K10}

\renewcommand{\thefootnote}{\arabic{footnote}}
\setcounter{footnote}{0}

\section{Introduction}

\subsection[Algebraic solutions of the Painlev\'e-III (D\_7) equation]{Algebraic solutions of the Painlev\'e-III ($\boldsymbol{{\rm D}_7}$) equation}

The Painlev\'e-III (D$_6$) equation for a function
$u\colon\mathbb{C}\to\mathbb{C}$, $x\mapsto u(x)$ is
\begin{equation*}
u'' = \frac{(u')^2}{u}-\frac{u'}{x}+\frac{\alpha u^2+\beta}{x}+\gamma u^3+\frac{\delta}{u},
\end{equation*}
where $\alpha,\beta,\gamma,\delta\in\mathbb{C}$ are parameters.
The Painlev\'e-III (D$_7$) equation
\begin{equation*}
u'' = \frac{(u')^2}{u}-\frac{u'}{x}+\frac{\alpha u^2+\beta}{x}+\frac{\delta}{u}
\end{equation*}
is a special degenerate case in which $\gamma=0$ and $\alpha\delta\neq 0$.
For more information on this equation see, for example, Kitaev and Vartanian
\cite{KitaevV:2004}.
With the choice of parameters $\alpha=8$, $\beta=2n$, and $\delta=-1$, namely
\begin{equation}
\label{p3d7}
u'' = \frac{(u')^2}{u}-\frac{u'}{x}+\frac{8u^2+2n}{x}-\frac{1}{u},
\end{equation}
the Painlev\'e-III (D$_7$) equation admits an algebraic solution for each
$n\in\mathbb{Z}$.
Specifically, define functions $R_n(\zeta)$ for $n\in\mathbb{Z}$ via
the recurrence relation
\begin{gather*}
 2\zeta R_{n+1}(\zeta)R_{n-1}(\zeta) = -R_n(\zeta)R_n''(\zeta)+R_n'(\zeta)^2
-\frac{1}{\zeta}R_n(\zeta)R_n'(\zeta)+2\bigl(\zeta^2-n\bigr)R_n(\zeta)^2, \\
 R_0(\zeta):=1, \qquad R_1(\zeta):=\zeta^2.
\end{gather*}
Examples of these functions are
\begin{alignat*}{4}
& R_{-3}(\zeta) = 1 + \frac{4}{\zeta^2} + \frac{5}{\zeta^4}, \qquad &&
R_{-2}(\zeta) = \frac{1}{\zeta} + \frac{1}{\zeta^3}, \qquad &&
R_{-1}(\zeta) = \frac{1}{\zeta},& \\
& R_2(\zeta) = \zeta^5-\zeta^3, \qquad &&
R_3(\zeta) = \zeta^9 - 4\zeta^7 + 5\zeta^5. \qquad &
\end{alignat*}
If $n\geq 0$, then $R_n(\zeta)$ is a polynomial in $\zeta$ known as an
\emph{Ohyama polynomial}~\cite{Ohyama:2006}. The unique \big(on the Riemann surface of $x^{1/3}$\big) algebraic solution to
\eqref{p3d7} is $u(x)=u_n(x)$, where
\begin{equation*}
u_n(x) := \frac{R_{n+1}\bigl(\sqrt{3}x^{1/3}\bigr)R_{n-1}\bigl(\sqrt{3}x^{1/3}\bigr)}{2\sqrt{3}R_n\bigl(\sqrt{3}x^{1/3}\bigr)^2},
\qquad n\in\mathbb{Z}.
\end{equation*}
The $u_n(x)$ are rational functions of $x^{1/3}$. If one selects the principal
branch for $x^{1/3}$, then each of these produces three distinct algebraic solutions
on the complex plane: $u_n(x)$ and $u_n\bigl(\mathrm{e}^{\pm 2\pi\mathrm{i}}x\bigr)$. Some
examples are
\begin{alignat*}{4}
& u_{-2}(x) = \frac{9x^{5/3}+12x+5x^{1/3}}{2\bigl(3x^{2/3}+1\bigr)^2}, \qquad &&
u_{-1}(x) = \frac{3x^{2/3}+1}{6x^{1/3}}, \qquad &&
u_0(x) = \frac{1}{2}x^{1/3},& \\
& u_1(x) = \frac{3x^{2/3}-1}{6x^{1/3}}, \qquad &&
u_2(x) = \frac{9x^{5/3}-12x+5x^{1/3}}{2\bigl(3x^{2/3}-1\bigr)^2}. \qquad &
\end{alignat*}
See Clarkson~\cite{Clarkson:2003} for additional background on these functions. The Painlev\'e-III (D$_7$) equation~\eqref{p3d7} is invariant under the symmetries $u(x)\mapsto \pm\mathrm{i} u(\pm\mathrm{i} x)$, $n\mapsto -n$, and it is easily seen that $\pm\mathrm{i} u_n(\pm\mathrm{i} x)=u_{-n}\bigl(\mathrm{e}^{\pm 2\pi\mathrm{i}} x\bigr)$.
In this paper, we will assume that $n\ge 0$ and also restrict attention to the principal sheet $-\pi<\arg(x)<\pi$.

\begin{figure}[t]\centering
\includegraphics[width=0.4\linewidth]{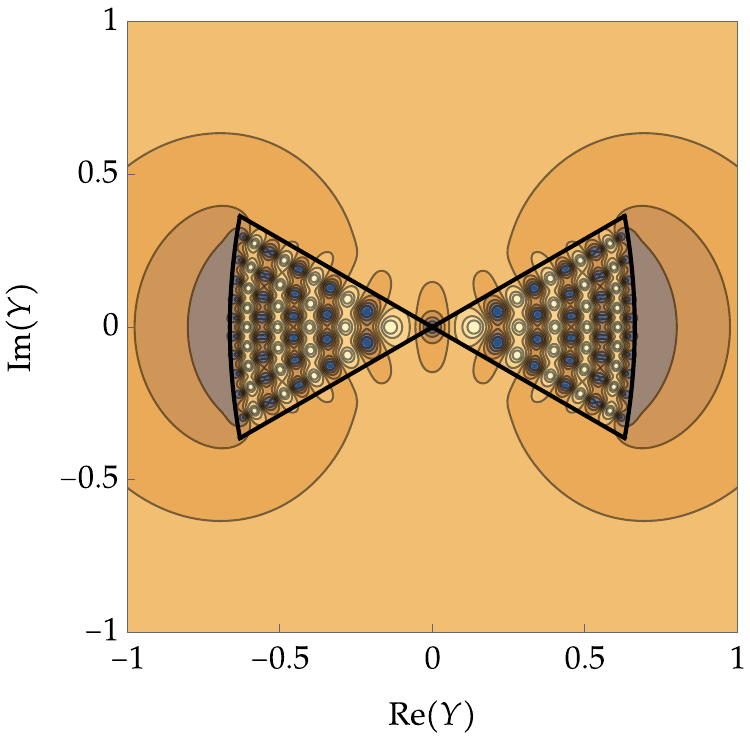}\hspace{0.1\linewidth}%
\includegraphics[width=0.4\linewidth]{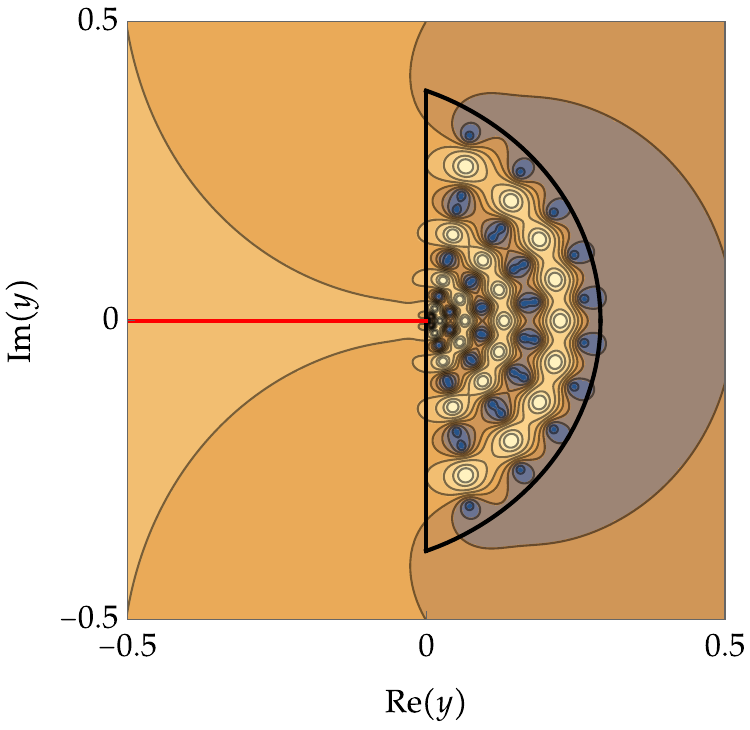}
\caption{Left panel: a density plot of $\big|U_{10}\bigl(Y^3\bigr)\big|$ and the boundary of the ``bow-tie'' region $\mathcal{B}$. Right panel: a similar plot of $|U_{10}(y)|$ on the principal sheet of the $y$-plane with $-\pi<\arg(y)<\pi$ and the sheet boundary (branch cut) shown with a red line. In both plots, lighter/darker color indicates larger/smaller modulus.}
\label{fig:DensityPlots}
\end{figure}

It is natural to introduce a scaled independent variable $y$ via
\begin{equation}
x=n^{3/2}y.
\label{eq:basicscaling}
\end{equation}
Under this scaling, plots show that the zeros and poles of $u_n\bigl(n^{3/2}y\bigr)$ appear to be confined for $n$ large to a ``bow-tie'' shaped bounded region in the $Y$-plane with $Y:=y^{1/3}$ that is asymptotically independent of $n$. See Figure~\ref{fig:DensityPlots}.
In~\cite{BuckinghamM22}, the limiting region was characterized precisely and it was proved that on the unbounded exterior of this region, the related function $U_n(y):=n^{-1/2}u_n\bigl(n^{3/2}y\bigr)$ converges as $n\to\infty$ to the solution $\breve{U}(y)$ of the cubic equation
\begin{equation}
8\breve{U}^3+2\breve{U}-y=0
\label{eq:equilibrium}
\end{equation}
that behaves as $\breve{U}\approx\frac{1}{2}y^{1/3}$ for large $y$ with $|\arg(y)|\le 3\pi$.

\subsection[Formal degeneration of Painlev\'e-III (D\_7)]{Formal degeneration of Painlev\'e-III (D$\boldsymbol{_7}$)}
\label{sec:scaling}
Motivated by this convergence result, we may refine the scaling~\eqref{eq:basicscaling} by introducing a second parameter $z\in\mathbb{C}$ and considering
\begin{equation}
U(z)=U_n\bigl(y+n^{-1}z\bigr)=n^{-1/2}u_n\bigl(n^{3/2}\bigl(y+n^{-1}z\bigr)\bigr)
\label{eq:U-of-z}
\end{equation}
as a function of $z$ for fixed $y\in\mathbb{C}$. Then, the Painlev\'e-III (D$_7$) equation~\eqref{p3d7} with parameter $n$ for $u_n(x)$ implies
\begin{equation*}
U''(z)=\frac{U'(z)^2}{U(z)}+\frac{8}{y}U(z)^2+\frac{2}{y}-\frac{1}{U(z)} +\mathcal{O}\bigl(n^{-1}\bigr),\qquad n\to\infty.
\end{equation*}
Thus, neglecting the $\mathcal{O}\bigl(n^{-1}\bigr)$ error term, the Painlev\'e-III (D$_7$) equation formally degenerates to a one-parameter family, parametrized by $y\in\mathbb{C}$, of autonomous differential equations that we write for an unknown $\breve{U}(z)=\breve{U}(z;y)$:
\begin{equation}
\breve{U}''(z)=\frac{\breve{U}'(z)^2}{\breve{U}(z)}+\frac{8}{y}\breve{U}(z)^2+\frac{2}{y}-\frac{1}{\breve{U}(z)}.
\label{eq:breveU-ODE}
\end{equation}
The cubic equation~\eqref{eq:equilibrium} corresponds to the solutions of~\eqref{eq:breveU-ODE} that are independent of $z$. For non-constant solutions,~\eqref{eq:breveU-ODE} can be multiplied by the integrating factor $\breve{U}'(z)/\breve{U}(z)^2$ and then integrated once to obtain
\begin{equation}
\breve{U}'(z)^2=\frac{16}{y}\breve{U}(z)^3+2E\breve{U}(z)^2-\frac{4}{y}\breve{U}(z)+1,
\label{eq:breveU-first-order}
\end{equation}
wherein $E\in\mathbb{C}$ is an integration constant. Setting $\breve{U}(z)=\frac{1}{4}y\wp(z-z_0)-\frac{1}{24}yE$ for arbitrary $z_0$ one finds that $\wp(z)$ solves the Weierstra\ss\ equation~\cite[Chapter 23]{DLMF}
\begin{equation}
\wp'(z)^2=4\wp(z)^3-g_2\wp(z)-g_3
\label{eq:Weierstrass}
\end{equation}
with invariants
\begin{equation*}
g_2=\frac{16}{y^2}+\frac{E^2}{3}\qquad\text{and}\qquad
g_3=-\frac{16}{y^2}-\frac{8E}{3y^2}-\frac{E^3}{27}.
\end{equation*}
Thus, one might expect that the algebraic solutions might be locally approximated near a point~$y$ with~$y^{1/3}$ in the bounded ``bow-tie'' region by a Weierstra\ss\ elliptic function of $z$ with invariants depending on~$y$ and~$E$. However, this formalism does not explain how $E$ should be chosen given~$y$, nor does it determine the offset $z_0$, and it is not a rigorous argument. For that one can use a Riemann--Hilbert characterization of $U_n(y)$ that was also found in~\cite{BuckinghamM22}. We describe this characterization next.

\subsection[Riemann--Hilbert representation of U\_n(y):=n\^{}\{-1/2\}u\_n(n\^{}\{3/2\}y)]{Riemann--Hilbert representation of $\boldsymbol{U_n(y):=n^{-1/2}u_n(n^{3/2}y)}$}
\begin{figure}[t]\centering
\includegraphics[width=0.23\linewidth]{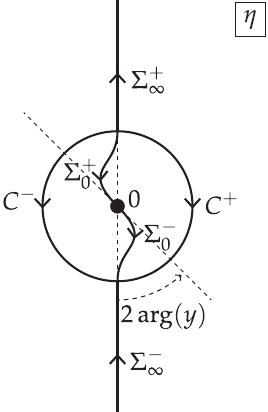}
\caption{The jump contour for Riemann--Hilbert Problem~\ref{rhp:Z}.}
\label{fig:Z-contour}
\end{figure}

Given $y\in\mathbb{C}$ with $|\arg(y)|<\pi$, let $\Sigma_0^\pm$, $\Sigma_\infty^\pm$, and $C^\pm$ be smooth, pairwise disjoint, oriented open arcs in the $\eta$-plane as shown in Figure~\ref{fig:Z-contour}.
The important properties of these arcs are the following.
\begin{itemize}\itemsep=0pt
\item $\Sigma_0^+$ terminates at the origin and $\Sigma_0^-$ originates from the origin tangent to the line making the angle $2\arg(y)$ with the vertical.
\item $\Sigma_\infty^+$ terminates vertically at $\eta=\mathrm{i}\infty$ and $\Sigma_\infty^-$ originates vertically from $\eta=-\mathrm{i}\infty$.
\item $\Sigma_\infty^+$, $\Sigma_0^+$, $C^+$, and $C^-$ share a common initial point.
\item $\Sigma_\infty^-$, $\Sigma_0^-$, $C^+$, and $C^-$ share a common terminal point.
\item $C^+$ (resp.\ $C^-$) lies in the component of $\mathbb{C}\setminus \overline{\Sigma_\infty^-\cup\Sigma_0^-\cup\Sigma_0^+\cup\Sigma_\infty^+}$ containing large positive (resp.\ negative) real values of $\eta$.
\end{itemize}
Denote
\begin{equation}
\Phi(\eta,y):=\mathrm{i}\eta-y(-\mathrm{i}\eta)^{-1/2}
\label{eq:Phi-def}
\end{equation}
with the branch cut of the function $\eta\mapsto\Phi(\eta,y)$ taken to coincide with $\overline{\Sigma_0^-\cup\Sigma_\infty^-}$ and $(-\mathrm{i}\eta)^{-1/2}$ taken to be real and positive for positive imaginary $\eta$ sufficiently large, and set
\begin{equation}
\widetilde{\mathbf{E}}:=\frac{1}{\sqrt{2}}\begin{bmatrix}\mathrm{e}^{5\pi\mathrm{i}/6} & \mathrm{e}^{-5\pi\mathrm{i}/6}\\\mathrm{e}^{-\mathrm{i}\pi/6} & \mathrm{e}^{-5\pi\mathrm{i}/6}\end{bmatrix}.
\label{eq:tilde-E}
\end{equation}
Consider the following problem, in which
$\sigma_3:=\mathrm{diag}(1,-1)$ denotes the third Pauli matrix.

\begin{rhp}[{scaled algebraic Painlev\'e-III (D$_7$) solutions,~\cite[Section~4.1]{BuckinghamM22}}]\label{rhp:Z}
Let $y\in\mathbb{C}$ with $|\arg(y)|\le\pi$ and $n\in\mathbb{Z}_{>0}$ be given. Seek a $2\times 2$ matrix function $\eta\mapsto\mathbf{Z}^{(n)}(\eta,y)$ with the following properties:
\begin{itemize}\itemsep=0pt
\item[]Analyticity: $\eta\mapsto\mathbf{Z}^{(n)}(\eta,y)$ is analytic for $\eta\in\mathbb{C}\setminus \overline{\Sigma_\infty^+\cup\Sigma_\infty^-\cup\Sigma_0^-\cup\Sigma_0^+\cup C^+\cup C^-}$.
\item[]Jump conditions: $\eta\mapsto \mathbf{Z}^{(n)}(\eta,y)$ takes continuous boundary values on the jump contour except at $\eta=0$, and these boundary values are related by the jump conditions
\begin{gather*}
\mathbf{Z}_+^{(n)}(\eta,y)=\mathbf{Z}^{(n)}_-(\eta,y)\mathrm{e}^{-n\Phi(\eta,y)\sigma_3}\begin{bmatrix}1&0\\\mathrm{i} & 1\end{bmatrix}\mathrm{e}^{n\Phi(\eta,y)\sigma_3},\qquad\eta\in\Sigma_\infty^+\cup\Sigma_0^+,
\\
\mathbf{Z}_+^{(n)}(\eta,y)=\mathbf{Z}^{(n)}_-(\eta,y)\mathrm{e}^{-n\Phi(\eta,y)\sigma_3}\begin{bmatrix}1&0\\-\mathrm{i} & 1\end{bmatrix}\mathrm{e}^{n\Phi(\eta,y)\sigma_3},\qquad\eta\in C^+\cup C^-,
\\
\mathbf{Z}_+^{(n)}(\eta,y)=\mathbf{Z}^{(n)}_-(\eta,y)\mathrm{e}^{-n\Phi_-(\eta,y)\sigma_3}(-1)^n\begin{bmatrix}1&-\mathrm{i} \\ 0 & 1\end{bmatrix}\mathrm{e}^{n\Phi_+(\eta,y)\sigma_3},\qquad\eta\in\Sigma_\infty^-,
\\
\mathbf{Z}_+^{(n)}(\eta,y)=\mathbf{Z}^{(n)}_-(\eta,y)\mathrm{e}^{-n\Phi_-(\eta,y)\sigma_3}\begin{bmatrix}0 & (-1)^n\mathrm{i}\\ (-1)^n\mathrm{i} & 0\end{bmatrix}
\mathrm{e}^{n\Phi_+(\eta,y)\sigma_3},\qquad\eta\in\Sigma_0^-.
\end{gather*}
Here a subscript $+$ $($resp.~$-)$ refers to a
boundary value taken on an oriented arc from its left $($resp.\ right$)$ side.
\item[]Normalization: $\mathbf{Z}^{(n)}(\eta,y)(-\mathrm{i}\eta)^{n\sigma_3/2}\to\mathbb{I}$ as $\eta\to\infty$.
\item[]Behavior as $\eta\to 0$: The limit of $\mathbf{Z}^{(n)}(\eta,y)\mathrm{e}^{-\mathrm{i} n\eta\sigma_3}\widetilde{\mathbf{E}}\cdot(-\mathrm{i}\eta)^{-(-1)^n\sigma_3/4}$ as $\eta\to 0$ exists.
\end{itemize}
In the last two conditions, powers of $-\mathrm{i}\eta$ are taken to be cut on $\Sigma_0^-\cup\Sigma_\infty^-$ and agree with principal branches for large positive imaginary $\eta$.
\end{rhp}

The last property can be used to define a matrix $\mathbf{B}^{(n)}_0(y)$ by
\begin{equation}
\mathbf{B}^{(n)}_0(y):=(-\mathrm{i} ny)^{n\sigma_3}\Big[\lim_{\eta\to 0}\mathbf{Z}^{(n)}(\eta,y)\mathrm{e}^{-\mathrm{i} n\eta\sigma_3}\widetilde{\mathbf{E}}\cdot(-\mathrm{i}\eta)^{-(-1)^n\sigma_3/4}\Big](-\mathrm{i} n y)^{(-1)^n\sigma_3/2},
\label{eq:B0n}
\end{equation}
with fractional powers of $-\mathrm{i} n y$ defined by continuation of the principal branch for $y>0$.
Then, a rescaling of the algebraic solution of Painlev\'e-III (D$_7$) is encoded in the solution of Riemann--Hilbert Problem~\ref{rhp:Z} by the formula
\begin{equation}
U_n(y):=n^{-1/2}u_n\bigl(n^{3/2}y\bigr) =ny\begin{cases}\mathrm{e}^{-5\pi\mathrm{i}/6}B_{0,12}^{(n)}(y)B_{0,22}^{(n)}(y),& \text{$n$ even}, \\
\mathrm{e}^{5\pi\mathrm{i}/6}B_{0,11}^{(n)}(y)B_{0,21}^{(n)}(y),& \text{$n$ odd.}
\end{cases}
\label{eq:Uny}
\end{equation}

\subsection{Main aims of the paper}
The conditions of Riemann--Hilbert Problem~\ref{rhp:Z} involve the large parameter $n$ in an explicit way and there are well-known techniques originating in the Deift--Zhou steepest-descent method~\cite{DeiftZ93} for analyzing such problems. One needs to firstly control the large exponential factors in the jump matrices by introducing an appropriate scalar $g$-function. The difference between~$g$ and~$\Phi$ is a function $h$ whose derivative satisfies an algebraic equation defining the \emph{spectral curve}. We show below that when $y$ corresponds to a point in the ``bow-tie'', the spectral curve has genus $1$ and that the landscape of $\operatorname{Re}(h)$ has the properties necessary to continue the analysis. The next step involves exploiting analytic factorizations of jump matrices to ``open lenses'' by moving certain factors off the jump contour onto nearby arcs. After this step, all jump matrices decay rapidly to the identity as $n\to\infty$ except along certain arcs where in the same limit a~nontrivial limiting jump matrix emerges instead. In the third step one uses the limiting jump matrix to define a Riemann--Hilbert problem for an approximation called an \emph{outer parametrix}; in addition one or more \emph{inner parametrices} are needed near certain points where the convergence of the jump matrix is not uniform. One pieces together a \emph{global parametrix} from the outer and inner parametrices to define an unjustified (at this point) approximation of the solution of the ``opened lenses'' problem. Finally, one proves a convergence theorem by showing that the matrix ratio of the unknown and its global parametrix solves a special kind of Riemann--Hilbert problem (a~\emph{small-norm problem}) for which the solution is uniformly close to the identity.

In this scheme, the approximate formula for $U_n(y)$ comes from the outer parametrix.
In the situation we discuss in this paper, that $y$ corresponds to a point in the
``bow-tie'', this outer parametrix can be written explicitly in terms of theta
functions of genus $1$ and elliptic integrals. Actually, we first replace $y$ with
$y+n^{-1}z$ as in~\eqref{eq:U-of-z} but use a $g$-function depending on $y$ only, and
then one obtains an approximate formula for $U_n\bigl(y+n^{-1}z\bigr)$ explicitly involving the
independent variable $z$ that should be related to the Weierstra\ss\ equation if the
formal reasoning described in Section~\ref{sec:scaling} above is correct. However,
it is very difficult to prove such a connection directly from the approximation
formula for $U_n\bigl(y+n^{-1}z\bigr)$; at the very least it is a calculation that is a~complicated
diversion from what should be a relatively simple path from~\eqref{p3d7}
to~\eqref{eq:breveU-first-order} or~\eqref{eq:Weierstrass}.\looseness=-1

Our aim in this paper is not to give all details of the convergence proof but rather
we focus on explaining a reasonably effective way to make the connection between the
outer parametrix Riemann--Hilbert problem -- whose conditions are far simpler than
the elliptic solution they generate -- and the limiting differential equation~\eqref{eq:breveU-first-order}. Our approach also determines the value of the
integration constant $E$ in~\eqref{eq:breveU-first-order} as a function of~$y$
(equivalently the value of both invariants in the Weierstra\ss\ equation
\eqref{eq:Weierstrass} are so-determined).
For those who would like to see the basic idea of this method illustrated
in a simple setting, in Appendix~\ref{app:toy-problem}, we show directly (without
reference to the known exact solution) that certain quantities derived from
a toy Riemann--Hilbert problem satisfy simple differential equations.
We originally learned this method from
Alexander Its (see, for example,~\cite[Chapter~8]{FokasIKN:2006} and
\cite{ItsK:2002}),
and it is a pleasure to write this article in his honor.\looseness=-1

\section{Spectral curves of genus 1}
Motivated by~\cite[Section 4.3]{BuckinghamM22}, for given complex parameters $y$ and $c$, we introduce a function $\eta\mapsto h_\eta(\eta,y,c)$ determined up to a sign by the equation
\begin{equation}
\left(\frac{\partial h}{\partial \eta}(\eta,y,c)\right)^2=f(\eta,y,c):=\frac{P(-\mathrm{i}\eta,y,c)}{(-\mathrm{i}\eta)^3},\qquad P(\mu,y,c):=-\mu^3+\mu^2+c\mu-\frac{y^2}{4}.
\label{eq:h-eta}
\end{equation}
Considered as an algebraic relation between $h_\eta$ and $\eta$, this defines the \emph{spectral curve}, which will have genus 1 provided $c$ and $y$ are chosen so that the three roots of the cubic $\mu\mapsto P(\mu,y,c)$ are distinct and nonzero. Note that if $y\neq 0$, $P(0,y,c)\neq 0$ so $\mu=0$ cannot be a root. Let us label the three distinct nonzero (for $y\neq 0$ and generic $c$) roots by $\mu=s_j$, $j=1,2,3$, so that $P(\mu,y,c)=-(\mu-s_1)(\mu-s_2)(\mu-s_3)$. Let $\Sigma_{0,1}$ be an arc in the $\eta$-plane joining $\eta=0$ to $\eta=\mathrm{i} s_1$, let $\Sigma_{2,3}$ be an arc in the $\eta$-plane joining $\eta=\mathrm{i} s_2$ to $\eta=\mathrm{i} s_3$, and assume that $\Sigma_{0,1}\cap\Sigma_{2,3}=\varnothing$.
Then we may define $h_\eta(\eta,y,c)$ unambiguously using~\eqref{eq:h-eta} by assuming that $\eta\mapsto h_\eta(\eta,y,c)$ is analytic for $\eta\in\mathbb{C}\setminus\overline{\Sigma_{0,1}\cup\Sigma_{2,3}}$ and that $h_\eta(\eta,y,c)=-\mathrm{i} -\frac{1}{2}\eta^{-1} +\mathcal{O}\bigl(\eta^{-2}\bigr)$ as $\eta\to\infty$. Then also,
\begin{equation*}
h_\eta(\eta,y,c)=\frac{1}{2}\mathrm{i} y(-\mathrm{i}\eta)^{-3/2}+O\bigl(\eta^{-1/2}\bigr),\qquad \eta\to 0,
\label{eq:hprime-zero}
\end{equation*}
with the power function being cut on $\Sigma_\infty^-\cup\Sigma_{0,1}$ and coinciding with the principal branch for large positive imaginary $\eta$.

Next we attempt to determine $c$ given $y\neq 0$ by imposing two real \emph{Boutroux conditions}:
\begin{gather}
I_{1,2}(y,c):=\operatorname{Re}\bigg(\int_{\mathrm{i} s_1}^{\mathrm{i} s_2}\frac{\partial h}{\partial \eta}(\eta,y,c)\,\mathrm{d}\eta\bigg)=0,\nonumber\\
I_{2,3}(y,c):=\operatorname{Re}\bigg(\int_{\mathrm{i} s_2}^{\mathrm{i} s_3}\frac{\partial h}{\partial\eta}(\eta,y,c)\,\mathrm{d}\eta\bigg)=0,
\label{eq:Boutroux}
\end{gather}
where the path of integration in each case lies in the domain of analyticity of $\eta\mapsto h_\eta(\eta,y,c)$. Although the latter domain is multiply connected, and hence $I_{1,2}$ and $I_{2,3}$ are only well-defined modulo a finitely-generated symmetry group, the conditions~\eqref{eq:Boutroux} are independent of the specific choice of paths due to the fact that $h_\eta(\eta,y,c)$ changes sign across its branch cuts and the fact that the residue of $h_\eta(\eta,y,c)$ at $\eta=\infty$ is real. If we introduce the real and imaginary parts of~$c$ by~$c_\mathrm{R}:=\operatorname{Re}(c)$ and $c_\mathrm{I}:=\operatorname{Im}(c)$ so that $c=c_\mathrm{R}+\mathrm{i} c_\mathrm{I}$, then we have
\begin{gather*}
\frac{\partial}{\partial c_\mathrm{R}}\bigg[\bigg(\frac{\partial h}{\partial\eta}(\eta,y,c)\bigg)^2\bigg] = -\eta^{-2},\\
\frac{\partial}{\partial c_\mathrm{I}}\bigg[\bigg(\frac{\partial h}{\partial\eta}(\eta,y,c)\bigg)^2\bigg] = -\mathrm{i}\eta^{-2},
\end{gather*}
from which it follows that if paths of integration are selected so that $I_{1,2}(y,c)$ and $I_{2,3}(y,c)$ depend smoothly on $(c_\mathrm{R},c_\mathrm{I})$,
\begin{equation*}
\begin{bmatrix}
\dfrac{\partial I_{1,2}}{\partial c_\mathrm{R}} & \dfrac{\partial I_{1,2}}{\partial c_\mathrm{I}}\vspace{1mm}\\
\dfrac{\partial I_{2,3}}{\partial c_\mathrm{R}} & \dfrac{\partial I_{2,3}}{\partial c_\mathrm{I}}
\end{bmatrix} =
\begin{bmatrix}
\displaystyle
-2\operatorname{Re}\bigg(\int_{\mathrm{i} s_1}^{\mathrm{i} s_2}\frac{\mathrm{d}\eta}{\eta^2h_\eta(\eta,y,c)}\bigg) & \displaystyle 2\operatorname{Im}\bigg(\int_{\mathrm{i} s_1}^{\mathrm{i} s_2}\frac{\mathrm{d}\eta}{\eta^2h_\eta(\eta,y,c)}\bigg)\vspace{1mm}\\
\displaystyle
-2\operatorname{Re}\bigg(\int_{\mathrm{i} s_2}^{\mathrm{i} s_3}\frac{\mathrm{d}\eta}{\eta^2h_\eta(\eta,y,c)}\bigg) & \displaystyle 2\operatorname{Im}\bigg(\int_{\mathrm{i} s_2}^{\mathrm{i} s_3}\frac{\mathrm{d}\eta}{\eta^2h_\eta(\eta,y,c)}\bigg)
\end{bmatrix}.
\end{equation*}
Therefore, the Jacobian determinant is
\begin{equation*}
\frac{\partial (I_{1,2},I_{2,3})}{\partial (c_\mathrm{R},c_\mathrm{I})}
= -4\operatorname{Im}\bigg(\bigg[\int_{\mathrm{i} s_1}^{\mathrm{i} s_2}\frac{\mathrm{d}\eta}{\eta^2h_\eta(\eta,y,c)}\bigg]^*
\int_{\mathrm{i} s_2}^{\mathrm{i} s_3}\frac{\mathrm{d}\eta}{\eta^2h_\eta(\eta,y,c)}\bigg).
\end{equation*}
Since $\bigl(\eta^2h_\eta(\eta,y,c)\bigr)^2 = -\mathrm{i}\eta P(-\mathrm{i}\eta,y,c)$ is a quartic polynomial with roots at $\eta=0$, $\mathrm{i} s_1$, $\mathrm{i} s_2$, $\mathrm{i} s_3$, the two integral factors on the right-hand side are complete elliptic integrals of the first kind over paths that form a basis for homology on the corresponding elliptic curve. Hence under the assumption that all four roots are distinct, the Jacobian is nonzero~\cite[Corollary 1]{Dubrovin81}. By the implicit function theorem, whenever a pair $y\neq 0$ and $c\in\mathbb{C}$ are such that both equations~\eqref{eq:Boutroux} hold and that $P(\cdot,y,c)$ has distinct roots, the solution $(c_\mathrm{R},c_\mathrm{I})$ of~\eqref{eq:Boutroux} can therefore be continued smoothly to nearby values of $y\in\mathbb{C}$.

We now assume that $y$ and $c$ are related so that the conditions~\eqref{eq:Boutroux} hold. The function $\eta\mapsto h_\eta(\eta,y,c)$ extends to a single-valued function on the two-sheeted Riemann surface $\mathcal{R}$ over the $\eta$-plane defined by the spectral curve~\eqref{eq:h-eta}. The differential $h_\eta(\eta,y,c)\,\mathrm{d}\eta$ is meromorphic on $\mathcal{R}$ with double poles (in suitable local coordinates) at the two points over $\eta=\infty$ and at the branch point $\eta=0$ and no other singularities. The residues of $h_\eta(\eta,y,c)\,\mathrm{d}\eta$ at the two points over $\eta=\infty$ are opposite real values $\pm\frac{1}{2}$ and then the residue at $\eta=0$ necessarily vanishes. It follows that if $c$ is such that the conditions~\eqref{eq:Boutroux} hold, then the multi-valued function $h(\eta,y,c)$ defined on $\mathcal{R}$ up to an integration constant by contour integration of $h_\eta(\eta,y,c)\,\mathrm{d}\eta$ has a real part that is single-valued on $\mathcal{R}$. Selecting the integration constant such that $\operatorname{Re}(h(\eta,y,c))$ vanishes at any one of the points $\eta=\mathrm{i} s_j$, $j=1,2,3$ (and hence at all three of them by~\eqref{eq:Boutroux}), the projection of the zero level set of $\operatorname{Re}(h(\eta,y,c))$ to either sheet of $\mathcal{R}$ is the same set in the $\eta$-plane, which we denote by $K$.

It is known~\cite[Theorem 3]{BuckinghamM22} that for large $n$, $u_n\bigl(n^{3/2}Y^3\bigr)$ is pole- and zero-free for $Y$ on an unbounded domain $\mathcal{E}$ whose complement is a ``bow-tie'' shaped region $\mathcal{B}:=\mathbb{C}\setminus\mathcal{E}$ in the $Y$-plane that is symmetric with respect to reflection in the real and imaginary axes. The interior of $\mathcal{B}$ is the disjoint union of two ``wings'', one on either side of the imaginary axis. The wings are joined at the origin only, and they are bounded in part by the straight-line segments joining the pairs $Y=\pm \bigl(2^{1/3}/3^{1/2}\bigr)\mathrm{e}^{\mathrm{i}\pi/6}$ and $\pm \bigl(2^{1/3}/3^{1/2}\bigr)\mathrm{e}^{5\pi\mathrm{i}/6}$. The set $\mathcal{B}\cap\mathbb{R}$ consists of the interval $\bigl[-y_\mathrm{c}^{1/3},y_\mathrm{c}^{1/3}\bigr]$, where $y_\mathrm{c}\approx 0.29177$. See the left-hand panel of Figure~\ref{fig:DensityPlots}.

\begin{figure}[t]\centering
\includegraphics[width=\linewidth]{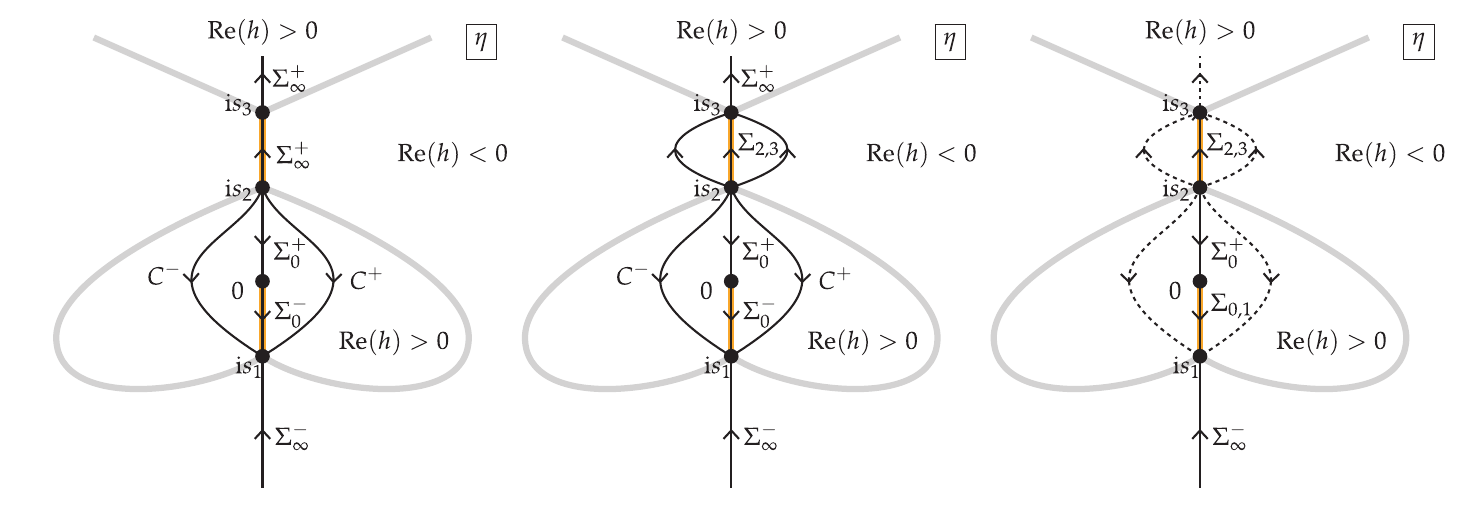}
\caption{Jump contours and sign chart of $\operatorname{Re}(h)$ in the $\eta$-plane for $0<y<y_\mathrm{c}$. Left panel: the zero level set $K$ of $\operatorname{Re}(h(\eta,y,c))$ shown in gray and orange (orange indicates the branch cuts~$\Sigma_{0,1}$ and~$\Sigma_{2,3}$ of~$h_\eta(\eta,y,c)$), and the relative placement of the jump contour for Riemann--Hilbert Problem~\ref{rhp:Z}. The sign of $\operatorname{Re}(h)$ is as indicated and $\operatorname{Re}(h)$ only changes sign across the gray arcs of $K$. Note that the contour~$\Sigma_\infty^+$ actually extends from $\eta=\mathrm{i} s_2$, taken as
the junction point of $C^-$, $C^+$, and $\Sigma_0^+$, all the way up the positive imaginary axis, passing through $\Sigma_{2,3}$. Likewise $\Sigma_\infty^-$ extends from $-\mathrm{i}\infty$ up to $\mathrm{i} s_1$. The arc $\Sigma_0^-$ coincides with the branch cut $\Sigma_{0,1}$. Center panel: the jump contour for $\mathbf{N}^{(n)}(\eta,y,z)$ has two additional arcs on the left and right of the branch cut $\Sigma_{2,3}$ after opening a lens. Right panel: the jump contour for $\breve{\mathbf{N}}^{(n),\mathrm{out}}(\eta,y,z)$ consists of the arcs $\Sigma_{\infty}^-$, $\Sigma_{0,1}$, $\Sigma_0^+$, and $\Sigma_{2,3}$ (shown with solid curves; the dashed arcs in the jump contour for $\mathbf{N}^{(n)}(\eta,y,z)$ have been neglected).}
\label{fig:InitialContours}
\end{figure}

\begin{Lemma}\label{lem:h}
Assume that $\operatorname{Re}(y)>0$ and that $Y=y^\frac{1}{3}$ lies in the open interior of $\mathcal{B}$. Then, there is a well-defined value $c=c_1(y)\in\mathbb{C}$, a smooth function of real variables $\operatorname{Re}(y)$ and $\operatorname{Im}(y)$ but not analytic in $y$, such that the
following hold.
\begin{itemize}\itemsep=0pt
\item $\eta\mapsto P(-\mathrm{i}\eta,y,c)$ has three distinct complex roots denoted $\eta=\mathrm{i} s_j$, $j=1,2,3$.
\item The Boutroux conditions~\eqref{eq:Boutroux} are satisfied.
\item There is a simple arc $\Sigma_h$ originating at $\eta=-\mathrm{i}\infty$ and terminating at one of the roots denoted $\eta=\mathrm{i} s_3$ that passes through in order the intermediate points $\eta=\mathrm{i} s_1$, $\eta=0$, and $\eta=\mathrm{i} s_2$, and an integration constant, such that $\eta\mapsto h(\eta,y,c)$ is analytic in $\mathbb{C}\setminus\Sigma_h$ and continuous up to $\Sigma_h\setminus\{0\}$
with
$\operatorname{Re}(h(\mathrm{i} s_j,y,c))=0$ for $j=1,2,3$, and such that $\eta\mapsto \operatorname{Re}(h(\eta,y,c))$ is continuous on $\mathbb{C}\setminus\{0\}$ and harmonic on $\mathbb{C}\setminus (\Sigma_{0,1}\cup\Sigma_{2,3})$, where $\Sigma_{0,1}$ denotes the arc of $\Sigma_h$ between $\eta=\mathrm{i} s_1$ and $\eta=0$ while $\Sigma_{2,3}$ denotes the arc of $\Sigma_h$ between $\eta=\mathrm{i} s_2$ and $\eta=\mathrm{i} s_3$, and the latter are taken to be the branch cuts of $\eta\mapsto h_\eta(\eta,y,c)$.
\item The zero level set $K$ of $\operatorname{Re}(h(\eta,y,c))$ consists of the arcs $\Sigma_{0,1}$ and $\Sigma_{2,3}$, two arcs joining $\eta=\mathrm{i} s_1$ to $\eta=\mathrm{i} s_2$ that bound a region containing $\eta=0$, and two unbounded arcs emanating from $\eta=\mathrm{i} s_3$, one tending to $\eta=\infty$ in the left half-plane and one tending to $\eta=\infty$ in the right half-plane. $\operatorname{Re}(h(\eta,y,c))$ changes sign across each of these arcs except for~$\Sigma_{0,1}$ and~$\Sigma_{2,3}$.
\item If $y>0$ \big(and also $y<y_\mathrm{c}\approx 0.29177$ so that $Y=y^{1/3}$ lies in $\mathcal{B}$\big) then $s_1<0<s_2<s_3$, $\Sigma_h$ consists of the part of the imaginary axis in the $\eta$-plane below the point $\eta=\mathrm{i} s_3$, and $K^*=-K$.
\end{itemize}
\end{Lemma}
We give the proof in the appendix. The structure of the set $K$ allows the arcs of the jump contour for Riemann--Hilbert Problem~\ref{rhp:Z} to be chosen in a useful way, as illustrated in the left-hand panel of Figure~\ref{fig:InitialContours} for $0<y<y_\mathrm{c}$.
The picture is topologically equivalent provided that $\operatorname{Re}(y)>0$ and $Y=y^{1/3}$ lies in the interior of $\mathcal{B}$ as in the conditions of Lemma~\ref{lem:h}.

\section[Introduction of g-function and lens opening]{Introduction of $\boldsymbol{g}$-function and lens opening}
We assume from now on that $\operatorname{Re}(y)>0$ and that $Y=y^{1/3}$ lies in the interior of $\mathcal{B}$.
Also, since $c=c_1(y)$ is determined from $y$ according to Lemma~\ref{lem:h}, we will write $h(\eta,y)=h(\eta,y,c_1(y))$ going forward. Under these assumptions, in this section we will implement the first two steps of the asymptotic analysis of Riemann--Hilbert Problem~\ref{rhp:Z} with $y$ replaced by $y+n^{-1}z$.

\subsection[First step: introduction of g-function]{First step: introduction of $\boldsymbol{g}$-function}
From $h(\eta,y)$, we define a related function by
\begin{equation*}
g=g(\eta,y)=\Phi(\eta,y)+h(\eta,y),\qquad \eta\in\mathbb{C}\setminus\Sigma_h,
\end{equation*}
with $\Phi(\eta,y)$ defined by~\eqref{eq:Phi-def}. In particular,
there is a function $g_0(y)$ such that
\begin{equation}
g(\eta,y)=-\frac{1}{2}\log(-\mathrm{i}\eta) + g_0(y)+\mathcal{O}\bigl(\eta^{-1/2}\bigr),\qquad\eta\to\infty.
\label{eq:g-large-eta}
\end{equation}
We then use $g(\eta,y)$ to modify the matrix $\mathbf{Z}^{(n)}\bigl(\eta,y+n^{-1}z\bigr)$ by setting
\begin{equation}
\mathbf{M}^{(n)}(\eta,y,z):=\mathrm{e}^{ng_0(y)\sigma_3}\mathbf{Z}^{(n)}\bigl(\eta,y+n^{-1}z\bigr)\mathrm{e}^{-ng(\eta,y)\sigma_3}.
\label{eq:MZ}
\end{equation}
Note that while $\mathbf{Z}^{(n)}\bigl(\eta,y+n^{-1}z\bigr)$ depends on $(y,z)$ only through the
combination $y+n^{-1}z$, the function $\mathbf{M}^{(n)}(\eta,y,z)$ involves these
variables in a more complicated fashion. However, as a~function of $\eta$,
$\mathbf{M}^{(n)}(\eta,y,z)$ is analytic where $\mathbf{Z}^{(n)}(\eta,y,z)$ is, and
according to~\eqref{eq:g-large-eta}, is normalized to the identity as
$\eta\to\infty$: $\mathbf{M}^{(n)}(\infty,y,z)=\mathbb{I}$.

\subsection{Second step: opening a lens}
We next open a lens about the branch cut $\Sigma_{2,3}$ of $h_\eta(\eta,y)$. Let $\Lambda^+$ (resp.\ $\Lambda^-$) denote a lens-shaped region abutting $\Sigma_{2,3}$ (oriented from $\mathrm{i} s_2(y)$ to $\mathrm{i} s_3(y)$) on its left (resp.\ right) side and lying in the domain where $\operatorname{Re}(h)<0$. We then define $\mathbf{N}^{(n)}(\eta,y,z)$ by setting
\begin{gather*}
\mathbf{N}^{(n)}(\eta,y,z):=\mathbf{M}^{(n)}(\eta,y,z)\begin{bmatrix}1&\mathrm{i}\mathrm{e}^{2nh(\eta,y)}\mathrm{e}^{2z(-\mathrm{i}\eta)^{-1/2}}\\0&1\end{bmatrix},\qquad \eta\in\Lambda^+,
\\
\mathbf{N}^{(n)}(\eta,y,z):=\mathbf{M}^{(n)}(\eta,y,z)\begin{bmatrix}1&-\mathrm{i}\mathrm{e}^{2nh(\eta,y)}\mathrm{e}^{2z(-\mathrm{i}\eta)^{-1/2}}\\0&1\end{bmatrix},\qquad\eta\in\Lambda^-,
\end{gather*}
and elsewhere we take $\mathbf{N}^{(n)}(\eta,y,z):=\mathbf{M}^{(n)}(\eta,y,z)$. The jumps of $\mathbf{N}^{(n)}(\eta,y,z)$ on the outer lens boundaries, oriented toward the endpoint $\eta=\mathrm{i} s_3$ as shown in the center panel of Figure~\ref{fig:InitialContours}, both read
\begin{equation*}
\mathbf{N}^{(n)}_+(\eta,y,z)=\mathbf{N}^{(n)}_-(\eta,y,z)\begin{bmatrix}1&-\mathrm{i}\mathrm{e}^{2nh(\eta,y)}\mathrm{e}^{2z(-\mathrm{i}\eta)^{-1/2}}\\0&1\end{bmatrix}.
\end{equation*}
On the part of $\Sigma_\infty^+$ coinciding with the branch cut $\Sigma_{2,3}$ in between the lenses, we get the modified jump condition
\begin{gather*}
\mathbf{N}^{(n)}_+(\eta,y,z) \\
\qquad
=\mathbf{N}^{(n)}_-(\eta,y,z)\begin{bmatrix}0 & \mathrm{i}\mathrm{e}^{n(h_+(\eta,y)+h_-(\eta,y))}\mathrm{e}^{2z(-\mathrm{i}\eta)^{-1/2}}\\\mathrm{i}\mathrm{e}^{-n(h_+(\eta,y)+h_-(\eta,y))}\mathrm{e}^{-2z(-\mathrm{i}\eta)^{-1/2}} & 0\end{bmatrix}.
\end{gather*}
By the Boutroux conditions~\eqref{eq:Boutroux} asserted in Lemma~\ref{lem:h}, it holds that $h_+(\eta,y)+h_-(\eta,y)=\mathrm{i}\psi(y)$ for $\eta\in\Sigma_{2,3}$, where $\psi(y)$ is a real quantity. By similar arguments as in~\cite{BuckinghamM22}, it also holds that $h_+(\eta,y)+h_-(\eta,y)=\mathrm{i}\xi(y)$ for $\eta\in\Sigma_0^-$, where $\xi(y)$ is a real quantity. Finally, $h_+(\eta,y)-h_-(\eta,y)=\mathrm{i}\kappa(y)$ for $\eta\in\Sigma_0^+=\Sigma_{0,1}$ where $\kappa(y)$ is a real quantity.

\section{Outer parametrix Riemann--Hilbert problem}

In addition to the conditions placed so far on $y$, we now suppose that $z\in\mathbb{C}$ is bounded. Then the jump matrices for $\mathbf{N}^{(n)}(\eta,y,z)$ all decay exponentially rapidly to the identity except near the two branch cuts $\Sigma_{0,1}$ and $\Sigma_{2,3}$ of $h_\eta(\eta,y)$, near the arc $\Sigma_{\infty}^-$, and near the arc $\Sigma_0^+$. In the branch-cut arcs, the jump conditions read as follows:
\begin{gather}
\mathbf{N}^{(n)}_+(\eta,y,z)=\mathbf{N}^{(n)}_-(\eta,y,z)\mathrm{e}^{z(-\mathrm{i}\eta)^{-1/2}\sigma_3}\!\!\begin{bmatrix}0 & \mathrm{i}\mathrm{e}^{\mathrm{i} n\psi(y)}\\\mathrm{i}\mathrm{e}^{-\mathrm{i} n\psi(y)} & 0\end{bmatrix}\!\mathrm{e}^{-z(-\mathrm{i}\eta)^{-1/2}\sigma_3}
\!\label{eq:Nn-jump-23}
\end{gather}
for $\eta\in\Sigma_{2.3}$, where the boundary values are determined by orientation of $\Sigma_{2,3}$ toward $\eta=\mathrm{i} s_3(y)$, and
\begin{gather}
 \mathbf{N}^{(n)}_+(\eta,y,z)=\mathbf{N}^{(n)}_-(\eta,y,z)\mathrm{e}^{z(-\mathrm{i}\eta)^{-1/2}_-\sigma_3}\begin{bmatrix}0 & (-1)^n\mathrm{i}\mathrm{e}^{\mathrm{i} n\xi(y)}\\
(-1)^n\mathrm{i}\mathrm{e}^{-\mathrm{i} n\xi(y)} & 0\end{bmatrix}\mathrm{e}^{-z(-\mathrm{i}\eta)^{-1/2}_+\sigma_3} \!\!\!
\label{eq:Nn-jump-10}
\end{gather}
for $\eta\in\Sigma_{0,1}$, where $\Sigma_{0,1}$ is oriented toward $\eta=\mathrm{i} s_1(y)$.
Note that the sum of the boundary values of~$(-\mathrm{i}\eta)^{-1/2}$ vanishes on this contour, so as the jump matrix is off-diagonal, the outer exponential factors in~\eqref{eq:Nn-jump-10} could have been omitted, but it is convenient to write them here anyway. The jump matrices in~\eqref{eq:Nn-jump-23} and~\eqref{eq:Nn-jump-10} are rapidly oscillatory in the parameter $y$, but the only dependence on $\eta$ enters via the diagonal conjugating factors with exponents proportional to $z$. Next, there is a residual jump across the contour $\Sigma_\infty^-$ with orientation toward $\eta=\mathrm{i} s_1(y)$ in the limit $n\to\infty$:
\begin{align*}
\mathbf{N}^{(n)}_+(\eta,y,z)&=\mathbf{N}^{(n)}_-(\eta,y,z)\mathrm{e}^{z(-\mathrm{i}\eta)^{-1/2}_-\sigma_3}\begin{bmatrix}1&-\mathrm{i}\mathrm{e}^{n(h_+(\eta,y)+h_-(\eta,y))}\\0 & 1\end{bmatrix}\mathrm{e}^{-z(-\mathrm{i}\eta)^{-1/2}_+\sigma_3}\\ &=
\mathbf{N}^{(n)}_-(\eta,y,z)\mathrm{e}^{z(-\mathrm{i}\eta)^{-1/2}_-\sigma_3}\left(\mathbb{I} + \text{exponentially small}\right)\mathrm{e}^{-z(-\mathrm{i}\eta)^{-1/2}_+\sigma_3}
\end{align*}
for $\eta\in\Sigma_\infty^-$, with the estimate of the error arising in the limit $n\to\infty$ because $\operatorname{Re}(h)<0$ on~$\Sigma_\infty^-$ (see Figure~\ref{fig:InitialContours}). Finally, there is also a residual jump across the contour $\Sigma_0^+$ with orientation toward $\eta=0$ in the limit $n\to\infty$:
\begin{align}
\mathbf{N}^{(n)}_+(\eta,y,z)&=\mathbf{N}^{(n)}_-(\eta,y,z)\mathrm{e}^{z(-\mathrm{i}\eta)^{-1/2}\sigma_3}\begin{bmatrix}
\mathrm{e}^{-\mathrm{i} n \kappa(y)} & 0 \\
\mathrm{i}\mathrm{e}^{-n(h_+(\eta,y)+h_-(\eta,y))} & \mathrm{e}^{\mathrm{i} n\kappa(y)}
\end{bmatrix}\mathrm{e}^{-z(-\mathrm{i}\eta)^{-1/2}\sigma_3}\label{eq:Nn-jump-Sigma0plus}
\\
&=\mathbf{N}^{(n)}_-(\eta,y,z)\mathrm{e}^{z(-\mathrm{i}\eta)^{-1/2}\sigma_3}\bigl(\mathrm{e}^{-\mathrm{i} n\kappa(y)\sigma_3} + \text{exponentially small}\bigr)\mathrm{e}^{-z(-\mathrm{i}\eta)^{-1/2}\sigma_3}\nonumber
\end{align}
for $\eta\in\Sigma_0^+$.

Neglecting the exponentially small terms,~\eqref{eq:Nn-jump-23}--\eqref{eq:Nn-jump-Sigma0plus} define the limiting jump conditions to be satisfied by an \emph{outer parametrix}. The convergence of the jump matrices overall to these three limits is not uniform near the three points $\eta=\mathrm{i} s_j(y)$, $j=1,2,3$, and one can
install standard \emph{inner parametrices} near each of these points constructed from Airy functions to correctly approximate $\mathbf{N}^{(n)}(\eta,y,z)$ nearby; see~\cite[Section 4.4.2]{BuckinghamM22} for some details in a very similar setting. However, no inner parametrix is needed near $\eta=0$ if one specifies suitable behavior for the outer parametrix at this point matching that inherited from $\mathbf{Z}^{(n)}\bigl(\eta,y+n^{-1}z\bigr)$.

By definition, the outer parametrix $\breve{\mathbf{N}}^{(n),\mathrm{out}}(\eta,y,z)$ is then the solution of the following Riemann--Hilbert problem, which retains all of the most important properties of $\mathbf{N}^{(n)}(\eta,y,z)$ when $\eta$ is bounded away from $\eta=\mathrm{i} s_j(y)$, $j=1,2,3$ and builds in the key property near these points needed to facilitate a good match between the outer and inner (Airy) parametrices near those points, namely a negative one-fourth power singularity.
\begin{rhp}\label{rhp:genus1outer}
Given $n\in\mathbb{Z}_{>0}$, $y\in\mathbb{C}$ with $\operatorname{Re}(y)>0$ and $Y=y^{1/3}$ in the interior of~$\mathcal{B}$, and $z\in\mathbb{C}$, seek a $2\times 2$ matrix-valued function $\bigl(\mathbb{C}\setminus \overline{\Sigma_{0,1}\cup\Sigma_0^+\cup\Sigma_{2,3}\cup\Sigma_\infty^-}\bigr)\ni\eta\mapsto\breve{\mathbf{N}}^{(n),\mathrm{out}}(\eta,y,z)$ with the following properties:
\begin{itemize}\itemsep=0pt
\item[] Analyticity: $\eta\mapsto\breve{\mathbf{N}}^{(n),\mathrm{out}}(\eta,y,z)$ is analytic in its domain of definition.
\item[]Jump conditions: $\eta\mapsto\breve{\mathbf{N}}^{(n),\mathrm{out}}(\eta,y,z)$ takes continuous boundary values from the left $(+$ subscript$)$ and right $(-$ subscript$)$ on the arc~$\Sigma_{0,1}$ oriented toward $\eta=\mathrm{i} s_1(y)$, the arc~$\Sigma_0^+$ oriented toward $\eta=0$, the arc~$\Sigma_{2,3}$ oriented toward $\eta=\mathrm{i} s_3(y)$, and the unbounded arc $\Sigma_{\infty}^-$ oriented toward $\eta=\mathrm{i} s_1(y)$, except near the finite endpoints $\eta=\mathrm{i} s_j(y)$, $j=1,2,3$. The jump conditions on these arcs are exactly the $n\to\infty$ limits of those satisfied by~$\mathbf{N}^{(n)}(\eta,y,z)$ $($see~\eqref{eq:Nn-jump-23}--\eqref{eq:Nn-jump-Sigma0plus}$)$:
\begin{gather}
 \breve{\mathbf{N}}^{(n),\mathrm{out}}_+(\eta,y,z)
=\breve{\mathbf{N}}^{(n),\mathrm{out}}_-(\eta,y,z)\mathrm{e}^{z(-\mathrm{i}\eta)^{-1/2}_-\sigma_3}\nonumber \\
\hphantom{\breve{\mathbf{N}}^{(n),\mathrm{out}}_+(\eta,y,z)=}{}\times \begin{bmatrix}0 & (-1)^n\mathrm{i}\mathrm{e}^{\mathrm{i} n\xi(y)} \nonumber \\
(-1)^n\mathrm{i}\mathrm{e}^{-\mathrm{i} n\xi(y)} & 0\end{bmatrix}\mathrm{e}^{-z(-\mathrm{i}\eta)^{-1/2}_+\sigma_3}, \qquad
 \eta\in\Sigma_{0,1}, \nonumber \\
 \breve{\mathbf{N}}^{(n),\mathrm{out}}_+(\eta,y,z)=\breve{\mathbf{N}}^{(n),\mathrm{out}}_-(\eta,y,z)\mathrm{e}^{z(-\mathrm{i}\eta)^{-1/2}\sigma_3}\mathrm{e}^{-\mathrm{i} n\kappa(y)\sigma_3}\mathrm{e}^{-z(-\mathrm{i}\eta)^{-1/2}\sigma_3},\qquad\eta\in\Sigma_0^+, \nonumber \\
 \breve{\mathbf{N}}^{(n),\mathrm{out}}_+(\eta,y,z) \nonumber \\
 \qquad=\breve{\mathbf{N}}^{(n),\mathrm{out}}_-(\eta,y,z)\mathrm{e}^{z(-\mathrm{i}\eta)^{-1/2}\sigma_3}\begin{bmatrix}0 & \mathrm{i}\mathrm{e}^{\mathrm{i} n\psi(y)}\\\mathrm{i}\mathrm{e}^{-\mathrm{i} n\psi(y)} & 0\end{bmatrix}\mathrm{e}^{-z(-\mathrm{i}\eta)^{-1/2}\sigma_3},\qquad \eta\in\Sigma_{2,3}, \nonumber \\
 \breve{\mathbf{N}}^{(n),\mathrm{out}}_+(\eta,y,z)=\breve{\mathbf{N}}^{(n),\mathrm{out}}_-(\eta,y,z)
\mathrm{e}^{z(-\mathrm{i}\eta)_-^{-1/2}\sigma_3}\mathrm{e}^{-z(-\mathrm{i}\eta)_+^{-1/2}\sigma_3},\qquad\eta\in\Sigma_\infty^-.
\label{eq:outer-parametrix-unbounded-jump}
\end{gather}
\item[]Normalization: $\breve{\mathbf{N}}^{(n),\mathrm{out}}(\eta,y,z)\to\mathbb{I}$ as $\eta\to\infty$.
\item[]Endpoint behavior: $\eta\mapsto\breve{\mathbf{N}}^{(n),\mathrm{out}}(\eta,y,z)$ is allowed to blow up like a negative one-fourth power near each of the four finite endpoints. In particular, there are matrices $\mathbf{A}^{(n)}_j=\mathbf{A}^{(n)}_j(y,z)$, $j=0,1,2,\dots$ independent of $\eta$ such that
\begin{gather}
\breve{\mathbf{N}}^{(n),\mathrm{out}}(\eta,y,z)\sim \!\Bigg(\sum_{j=0}^\infty \mathbf{A}_j(-\mathrm{i}\eta)^j\Bigg)\!(-\mathrm{i}\eta)^{(-1)^n\sigma_3/4}\widetilde{\mathbf{E}}^{-1}\mathrm{e}^{\mathrm{i} n\eta\sigma_3}\mathrm{e}^{-ng(\eta,y)\sigma_3},\quad\eta\to 0,
\!\label{eq:parametrix-at-0}
\end{gather}
where $\tilde{\mathbf{E}}$ is defined by~\eqref{eq:tilde-E}.
\end{itemize}
\end{rhp}

The jump contour for $\breve{\mathbf{N}}^{(n),\mathrm{out}}(\eta,y,z)$ is illustrated in the right-hand panel of Figure~\ref{fig:InitialContours}. This Riemann--Hilbert problem can be solved explicitly, but the construction is not as simple as the above conditions suggest. It involves elliptic integrals on the genus-1 spectral curve and corresponding Jacobi theta functions. Full details of the solution of a similar problem can be found in~\cite[Section 4.4.2]{BothnerM20}, for instance. The solution formula shows that, given~$n$ and~$y$, $\breve{\mathbf{N}}^{(n),\mathrm{out}}(\eta,y,z)$ exists for all $z\in\mathbb{C}$ except for a doubly periodic lattice of isolated points. However, we will have no need of the resulting complicated formul\ae\ in this paper.

Replacing $y$ with $y+n^{-1}z$ in~\eqref{eq:B0n} and \eqref{eq:Uny}, the rescaled algebraic solution $U_n\bigl(y+n^{-1}z\bigr)$ is expressed in terms of $\mathbf{Z}^{(n)}\bigl(\eta,y+n^{-1}z\bigr)$ by
\begin{equation*}
U_n\bigl(y+n^{-1}z\bigr) = (ny+z)\begin{cases}\mathrm{e}^{-5\pi\mathrm{i}/6}B^{(n)}_{0,12}\bigl(y+n^{-1}z\bigr)B^{(n)}_{0,22}\bigl(y+n^{-1}z\bigr),&\text{$n$ even},\\\mathrm{e}^{5\pi\mathrm{i}/6}B^{(n)}_{0,11}\bigl(y+n^{-1}z\bigr)B^{(n)}_{0,21}\bigl(y+n^{-1}z\bigr),&\text{$n$ odd},\end{cases}
\end{equation*}
wherein
\begin{gather*}
 \mathbf{B}_0^{(n)}\bigl(y+n^{-1}z\bigr)=(-\mathrm{i} (ny+z))^{n\sigma_3} \\
\qquad{}\times\Big[\lim_{\eta\to 0}\mathbf{Z}^{(n)}\bigl(\eta,y+n^{-1}z\bigr)\mathrm{e}^{-\mathrm{i} n\eta\sigma_3}\widetilde{\mathbf{E}}\cdot(-\mathrm{i} \eta)^{-(-1)^n\sigma_3/4}\Big](-\mathrm{i} (ny+z))^{(-1)^n\sigma_3/2}.
\end{gather*}
Also, writing $\mathbf{Z}^{(n)}\bigl(\eta,y+n^{-1}z\bigr)$ in terms of $\mathbf{M}^{(n)}(\eta,y,z)$ by~\eqref{eq:MZ} and using the fact that $\mathbf{M}^{(n)}(\eta,y,z)=\mathbf{N}^{(n)}(\eta,y,z)$ identically for $\eta$ in a neighborhood of the origin, 
\begin{gather*}
\lim_{\eta\to 0}\mathbf{Z}^{(n)}\bigl(\eta,y+n^{-1}z\bigr)\mathrm{e}^{-\mathrm{i} n\eta\sigma_3}\widetilde{\mathbf{E}}\cdot (-\mathrm{i}\eta)^{-(-1)^n\sigma_3/4} \\
\qquad = \mathrm{e}^{-ng_0(y)\sigma_3}\lim_{\eta\to 0}\mathbf{N}^{(n)}(\eta,y,z)\mathrm{e}^{ng(\eta,y)\sigma_3}\mathrm{e}^{-\mathrm{i} n\eta\sigma_3}\widetilde{\mathbf{E}}\cdot (-\mathrm{i}\eta)^{-(-1)^n\sigma_3/4}.
\end{gather*}
It then follows that
\begin{equation*}
U_n\bigl(y+n^{-1}z\bigr) = \begin{cases}\mathrm{i}\mathrm{e}^{-5\pi\mathrm{i}/6}\widetilde{B}^{(n)}_{0,12}(y,z)\widetilde{B}^{(n)}_{0,22}(y,z),&\text{$n$ even},\\\mathrm{i}\mathrm{e}^{5\pi\mathrm{i}/6}\widetilde{B}^{(n)}_{0,11}(y,z)\widetilde{B}^{(n)}_{0,21}(y,z),&\text{$n$ odd},\end{cases}
\end{equation*}
wherein
\begin{equation*}
\widetilde{\mathbf{B}}^{(n)}_0(y,z):=\lim_{\eta\to 0}\mathbf{N}^{(n)}(\eta,y,z)\mathrm{e}^{ng(\eta,y)\sigma_3}\widetilde{\mathbf{E}}\cdot (-\mathrm{i}\eta)^{-(-1)^n\sigma_3/4}.
\end{equation*}
To obtain an approximation $\breve{U}_n(z;y)$ of $U_n\bigl(y+n^{-1}z\bigr)$, we replace
the expression $\mathbf{N}^{(n)}(\eta,y,z)$ with $\breve{\mathbf{N}}^{(n),\mathrm{out}}(\eta,y,z)$ in this formula and, then using~\eqref{eq:parametrix-at-0} we get
\begin{equation}
\breve{U}_n(z;y):=\begin{cases}\mathrm{i}\mathrm{e}^{-5\pi\mathrm{i}/6}A^{(n)}_{0,12}(y,z)A^{(n)}_{0,22}(y,z),&\text{$n$ even},\\\mathrm{i}\mathrm{e}^{5\pi\mathrm{i}/6}A^{(n)}_{0,11}(y,z)A^{(n)}_{0,21}(y,z),&\text{$n$ odd}.\end{cases}
\label{eq:Un-approx}
\end{equation}
Note that by taking the limit $\eta\to 0$ from the left and right sides of the jump contour through the origin, one obtains two equivalent formul\ae\ for the matrix coefficient $\mathbf{A}^{(n)}_0(y,z)$:
\begin{equation*}
\mathbf{A}^{(n)}_0(y,z)=\lim_{\eta\to 0}\breve{\mathbf{N}}^{(n),\mathrm{out}}_\pm(\eta,y,z)\mathrm{e}^{ng_\pm(0,y)\sigma_3}\widetilde{\mathbf{E}}\cdot (-\mathrm{i}\eta)^{-(-1)^n\sigma_3/4},
\end{equation*}
where the $\pm$ signs correspond in the two instances.

Accuracy of the approximation of $U_n\bigl(y+n^{-1}z\bigr)$ by $\breve{U}_n(z;y)$ in the limit of large $n$ hinges on the details of the analysis of a small-norm Riemann--Hilbert problem for the matrix ratio between $\mathbf{N}^{(n)}(\eta,y,z)$ and its global parametrix. This is important, but it takes us far from our main goal in this work, which is to explain how one can prove, relatively easily and directly from the conditions of Riemann--Hilbert Problem~\ref{rhp:genus1outer}, that $\breve{U}(z)=\breve{U}_n(z;y)$ as defined in~\eqref{eq:Un-approx} is an exact solution of the elliptic differential equation~\eqref{eq:breveU-first-order} for a specific choice of the integration constant $E$ as a function of $y$.

\section[Derivation of the Weierstrass differential equation for breve{U}\_\{n\}(z;y)]{Derivation of the Weierstra\ss\ differential equation for $\boldsymbol{\breve{U}_n(z;y)}$}
It is a familiar outcome that various coefficients in the expansion of the solution of a Riemann--Hilbert problem depending on a parameter $z$ satisfy important differential equations. Indeed, this is exactly how one can be sure that Riemann--Hilbert Problem~\ref{rhp:Z} generates a rescaled solution of the Painlev\'e-III (D$_7$) equation~\eqref{p3d7} by formula~\eqref{eq:Uny} for each $n$. Such a computation is done in~\cite[Section 3.2]{BuckinghamM22} for a Riemann--Hilbert problem equivalent to Riemann--Hilbert Problem~\ref{rhp:Z} but with an unknown denoted $\mathbf{W}^{(n)}(\lambda,x)$. The steps are as follows:
\begin{itemize}\itemsep=0pt
\item One first introduces a diagonal exponential transformation by setting
\begin{equation*}
\boldsymbol{\Psi}^{(n)}(\lambda,x):=\mathbf{W}^{(n)}(\lambda,x)\mathrm{e}^{-\mathrm{i} (x\lambda-x(\mathrm{i} x\lambda)^{-1/2})\sigma_3}.
\end{equation*}
This has the effect of making the induced jump matrices for $\boldsymbol{\Psi}^{(n)}(\lambda,x)$ arcwise independent of both $\lambda$ (the complex variable of the Riemann--Hilbert problem) and $x$ (the independent variable of the Painlev\'e-III (D$_7$) equation in the form~\eqref{p3d7}).
\item It then follows by differentiation of the jump conditions that the matrices
\begin{gather}
\boldsymbol{\Lambda}^{(n)}(\lambda,x):=\frac{\partial\boldsymbol{\Psi}^{(n)}}{\partial\lambda}(\lambda,x)\boldsymbol{\Psi}^{(n)}(\lambda,x)^{-1},\nonumber\\
 \mathbf{X}^{(n)}(\lambda,x):=\frac{\partial\boldsymbol{\Psi}^{(n)}}{\partial x}(\lambda,x)\boldsymbol{\Psi}^{(n)}(\lambda,x)^{-1}
 \label{eq:LambdaX}
 \end{gather}
are analytic in $\lambda$ except at isolated singular points which in this case are $\lambda=\infty$ and $\lambda=0$.
\item By expanding $\boldsymbol{\Psi}^{(n)}(\lambda,x)$ and its derivatives near the singular points using information from the Riemann--Hilbert problem for $\mathbf{W}^{(n)}(\lambda,x)$, one deduces that both~$\boldsymbol{\Lambda}^{(n)}(\lambda,x)$ and~$\mathbf{X}^{(n)}(\lambda,x)$ are rational functions of $\lambda$ with principal parts expressed in terms of expansion coefficients of $\mathbf{W}^{(n)}(\lambda,x)$.
\item Re-arranging the equations~\eqref{eq:LambdaX} with this new knowledge, one sees that $\boldsymbol{\Psi}^{(n)}(\lambda,x)$ satisfies an overdetermined system consisting of two first-order $2\times 2$ linear systems, one with respect to $\lambda$ and another with respect to $x$.
\item Expressing the compatibility condition between the two systems in terms of the elements of the matrices $\boldsymbol{\Lambda}^{(n)}(\lambda,x)$ and $\mathbf{X}^{(n)}(\lambda,x)$, one separates out from the various powers of~$\lambda$ a~closed system of nonlinear differential equations on the coefficients with respect to $x$ alone. This system implies the Painlev\'e-III (D$_7$) equation~\eqref{p3d7}.
\end{itemize}
Analogues of these steps are frequently called the \emph{dressing method} in many papers.

It is a natural expectation that a similar approach might apply to Riemann--Hilbert Problem~\ref{rhp:genus1outer} to allow one to deduce a differential equation with respect to $z$ satisfied by $\breve{U}_n(z;y)$. Indeed, the matrix function
\begin{equation}
\eta\mapsto\mathbf{F}^{(n)}(\eta,y,z):=\breve{\mathbf{N}}^{(n),\mathrm{out}}(\eta,y,z)\mathrm{e}^{z(-\mathrm{i}\eta)^{-1/2}\sigma_3}
\label{eq:F-N}
\end{equation}
satisfies modified jump conditions that simply omit the factors $\mathrm{e}^{\pm z(-\mathrm{i}\eta)_{\mp}^{-1/2}\sigma_3}$ from the jump matrix. Hence the jump matrices for $\mathbf{F}^{(n)}(\eta,y,z)$ are arcwise independent of both $\eta$ and $z$. One can then derive a linear first-order differential equation for $\mathbf{F}^{(n)}(\eta,y,z)$ with respect to $z$ (see Section~\ref{sec:ODEz} below). However, derivation of a linear first-order differential equation for $\mathbf{F}^{(n)}(\eta,y,z)$ with respect to $\eta$ is more challenging. One can deduce that $\mathbf{F}^{(n)}_\eta(\eta,y,z)\mathbf{F}^{(n)}(\eta,y,z)^{-1}$ is rational in $\eta$ with simple poles at $\eta=0$ and $\eta=\mathrm{i} s_j(y)$, $j=1,2,3$, but it turns out that there is not enough information available to deduce fully the residue matrices. Without the first-order system with respect to $\eta$ one cannot obtain the desired nonlinear differential equation from any compatibility condition.

About a decade ago, we approached Alexander Its with a similar conundrum in the
setting of a project to study elliptic function approximations of rational solutions
of the second Painlev\'e equation~\cite{BuckinghamM14}. His advice was to
eschew the undetermined Fuchsian linear system with respect to the Riemann--Hilbert
complex variable (spectral parameter) in favor of a remarkable algebraic identity
satisfied by the matrix solutions of Riemann--Hilbert problems whose jump matrices
have a certain structure. Expanding this identity with respect to the spectral
parameter produces numerous identities among functions of the independent variable
alone that serve to close the system of differential equations; squaring it produces
a scalar identity that links the spectral curve and the target differential equation.

The jump matrices of Riemann--Hilbert Problem~\ref{rhp:genus1outer} have the necessary structure for this method to apply. In the rest of this section, we implement the method and show how it yields the expected differential equation~\eqref{eq:breveU-first-order}. Specifically, we prove the following.
\begin{Theorem}\label{thm:main-goal}
Fix $n\in\mathbb{Z}_{>0}$ and $y$ with $\operatorname{Re}(y)>0$ and $Y=y^{1/3}$ in the
interior of $\mathcal{B}$. Then the function $z\mapsto \breve{U}_n(z;y)$ defined
from the solution of Riemann--Hilbert Problem~$\ref{rhp:genus1outer}$ by
\eqref{eq:Un-approx} is a~solution of the differential equation
\eqref{eq:breveU-first-order}, in which the integration constant $E$ is given by
\begin{equation}
E=-\frac{8c}{y^2},
\label{eq:Ec}
\end{equation}
where $c=c_1(y)$ is the smooth but non-analytic function of $y$ defined in Lemma~$\ref{lem:h}$.
\end{Theorem}

\begin{Remark}
This result shows that the first-order autonomous differential equation
\eqref{eq:breveU-first-order}, which is now well-defined given $y$ as in the theorem
statement, is solved by the approximation~$\breve{U}_n(z;y)$, which also depends on
the index $n\in\mathbb{Z}_{>0}$. However, the space of solutions of the differential
equation is mapped out by translations in $z$, and the particular translate needed to
identify $\breve{U}_n(z;y)$ will generally depend on $n$ and is not specified by
Theorem~\ref{thm:main-goal}.
\end{Remark}

\begin{Remark}
Theorem~\ref{thm:main-goal} shows that the (scaled) algebraic function
$n^{-1/2}u_n\bigl(n^{3/2}\bigl(y+n^{-1}z\bigr)\bigr)$, which has a finite number of poles, is well approximated in
its pole region as $n\to\infty$ by a solution of the Weierstra\ss\ equation
in the form~\eqref{eq:breveU-first-order} having an infinite number of poles. Interestingly, the same Weierstra\ss\ equation has
recently been shown to govern large-$x$ asymptotic behavior of general (non-algebraic)
solutions of the Painlev\'e-III (D$_7$) equation~\eqref{p3d7} by Shimomura
\cite{Shimomura:2022}.
\end{Remark}

Now we continue with the proof of Theorem~\ref{thm:main-goal}. An elementary example illustrating the basic steps in the method we use can be found in Appendix~\ref{app:toy-problem}.

\subsection[Expansion of breve\{N\}\^{}\{(n),out\}(eta,y,z) near eta = infty]{Expansion of $\boldsymbol{\breve{\mathbf{N}}^{(n),\mathrm{out}}(\eta,y,z)}$ near $\boldsymbol{\eta=\infty}$}

It is easy to see from the jump condition~\eqref{eq:outer-parametrix-unbounded-jump} that if $y$, $z$, and $n$ are such that $\breve{\mathbf{N}}^{(n),\mathrm{out}}(\eta,y,z)$ exists, the product $\mathbf{F}(\eta)=\mathbf{F}^{(n)}(\eta,y,z)$ defined by~\eqref{eq:F-N} is analytic for large $\eta$ and decays to $\mathbb{I}$ as $\eta\to\infty$. Therefore, there are matrix coefficients $\mathbf{F}_j=\mathbf{F}_j^{(n)}(y,z)$ such that
\begin{gather}
\breve{\mathbf{N}}^{(n),\mathrm{out}}(\eta,y,z)=\mathbf{F}(\eta)\mathrm{e}^{-z(-\mathrm{i}\eta)^{-1/2}\sigma_3}
 =\Bigg(\!\mathbb{I} + \sum_{j=1}^\infty\mathbf{F}_j\cdot(-\mathrm{i}\eta)^{-j}\!\Bigg)\sum_{k=0}^\infty\frac{(-z)^k}{k!}(-\mathrm{i}\eta)^{-k/2}\sigma_3^k \!\!\!\label{eq:OuterParametrix-H-Expansion}\\
\qquad=\Bigg(\mathbb{I} + \sum_{j=1}^\infty\mathbf{F}_j\cdot(-\mathrm{i}\eta)^{-j}\Bigg)\Bigg(\mathbb{I}\sum_{k=0}^\infty\frac{z^{2k}}{(2k)!}(-\mathrm{i}\eta)^{-k} - (-\mathrm{i}\eta)^{-1/2}\sigma_3\sum_{k=0}^\infty\frac{z^{2k+1}}{(2k+1)!}(-\mathrm{i}\eta)^{-k}\Bigg). \nonumber
\end{gather}
This immediately implies that $\breve{\mathbf{N}}^{(n),\mathrm{out}}(\eta,y,z)$ has an expansion in nonnegative integer powers of~$(-\mathrm{i}\eta)^{-1/2}$
\begin{equation}
\breve{\mathbf{N}}^{(n),\mathrm{out}}(\eta,y,z)=\sum_{m=0}^\infty \mathbf{C}_{m/2}\cdot (-\mathrm{i}\eta)^{-m/2},\qquad \mathbf{C}_0=\mathbb{I}
\label{eq:OuterParametrix-InfinityExpansion}
\end{equation}
that is convergent for $|\eta|$ large enough.

By matching the coefficients of like powers of $-\mathrm{i}\eta$ in the expansions~\eqref{eq:OuterParametrix-H-Expansion}--\eqref{eq:OuterParametrix-InfinityExpansion}, one can express the coefficients $\mathbf{C}_{m/2}=\mathbf{C}_{m/2}^{(n)}(y,z)$ in terms of the $\mathbf{F}_j$. Therefore, we find
\begin{gather}
 \mathbf{C}_{1/2} =-z\sigma_3,\qquad
 \mathbf{C}_1 = \mathbf{F}_1 + \frac{z^2}{2!}\mathbb{I},\qquad
 \mathbf{C}_{3/2} = -z\mathbf{F}_1\sigma_3 -\frac{z^3}{3!}\sigma_3,\label{eq:Cs-first}
\end{gather}
and so on. Eliminating $\mathbf{F}_1$ in favor of $\mathbf{C}_1$, we can also write
\begin{equation}
\mathbf{C}_{3/2}=-z\mathbf{C}_1\sigma_3 +\frac{z^3}{3}\sigma_3.
\label{eq:C3/2-C1}
\end{equation}

\begin{Remark}\label{rem:zero}
Similar analysis of the growth and jump conditions of Riemann--Hilbert Problem~\ref{rhp:genus1outer} near $\eta=0$ shows that the product $\breve{\mathbf{N}}^{(n),\mathrm{out}}(\eta,y,z)\mathrm{e}^{ng(\eta,y)\sigma_3}\mathrm{e}^{-\mathrm{i} n\eta\sigma_3}\widetilde{\mathbf{E}}\cdot(-\mathrm{i}\eta)^{-(-1)^n\sigma_3/4}$ is analytic at $\eta=0$. This implies that the series in~\eqref{eq:parametrix-at-0} is convergent, and that ``$\sim$'' can be written instead as ``$=$''.
\end{Remark}

\subsection[Differential equations in z]{Differential equations in $\boldsymbol{z}$}\label{sec:ODEz}

\subsubsection[Lax equation satisfied by F\^{}\{(n)\}(eta,y,z)]{Lax equation satisfied by $\boldsymbol{\mathbf{F}^{(n)}(\eta,y,z)}$}

Assuming it exists, the matrix function $\eta\mapsto \mathbf{F}^{(n)}(\eta,y,z)$ defined in~\eqref{eq:F-N} above satisfies modified jump conditions with jump matrices that are independent of $z$ (and also of $\eta$, but we will not use that). Then by standard arguments, $\mathbf{F}^{(n)}_z(\eta,y,z)\mathbf{F}^{(n)}(\eta,y,z)^{-1}$ is analytic in $\eta$ except possibly at $\eta=0,\infty$. In terms of the outer parametrix, we have from~\eqref{eq:F-N}
\begin{gather*}
\frac{\partial\mathbf{F}^{(n)}}{\partial z}(\eta,y,z)\mathbf{F}^{(n)}(\eta,y,z)^{-1} = \frac{\partial\breve{\mathbf{N}}^{(n),\mathrm{out}}}{\partial z}(\eta,y,z)\breve{\mathbf{N}}^{(n),\mathrm{out}}(\eta,y,z)^{-1} \\
\hphantom{\frac{\partial\mathbf{F}^{(n)}}{\partial z}(\eta,y,z)\mathbf{F}^{(n)}(\eta,y,z)^{-1} =}{}
+ (-\mathrm{i}\eta)^{-1/2}\breve{\mathbf{N}}^{(n),\mathrm{out}}(\eta,y,z)\sigma_3\breve{\mathbf{N}}^{(n),\mathrm{out}}(\eta,y,z)^{-1}.
\end{gather*}
The expansion~\eqref{eq:OuterParametrix-InfinityExpansion} is differentiable term-by-term with respect to $z$, and therefore
\begin{gather*}
\frac{\partial\breve{\mathbf{N}}^{(n),\mathrm{out}}}{\partial z} (\eta,y,z)\breve{\mathbf{N}}^{(n),\mathrm{out}}(\eta,y,z)^{-1}\\
 \qquad{}= \bigg(\frac{\partial\mathbf{C}_{1/2}}{\partial z} \cdot(-\mathrm{i}\eta)^{-1/2} +\frac{\partial\mathbf{C}_1}{\partial z}\cdot (-\mathrm{i}\eta)^{-1} + \cdots\bigg) \\
 \qquad\quad{} \times\bigl(\mathbb{I}-\mathbf{C}_{1/2}(-\mathrm{i}\eta)^{-1/2} + \big[\mathbf{C}_{1/2}^2-\mathbf{C}_1\big](-\mathrm{i}\eta)^{-1} + \cdots\bigr)\\
 \qquad{} = \frac{\partial \mathbf{C}_{1/2}}{\partial z}\cdot (-\mathrm{i}\eta)^{-1/2} + \mathcal{O}\bigl((-\mathrm{i}\eta)^{-1}\bigr)
 = -\sigma_3(-\mathrm{i}\eta)^{-1/2}+\mathcal{O}\bigl((-\mathrm{i}\eta)^{-1}\bigr),\qquad\eta\to\infty,
\end{gather*}
where we used $\mathbf{C}_{1/2}=-z\sigma_3$. Also, directly from~\eqref{eq:OuterParametrix-InfinityExpansion},
\begin{gather*}
(-\mathrm{i}\eta)^{-1/2}\breve{\mathbf{N}}^{(n),\mathrm{out}}(\eta,y,z)\sigma_3\breve{\mathbf{N}}^{(n),\mathrm{out}}(\eta,y,z)^{-1}=\sigma_3 (-\mathrm{i}\eta)^{-1/2} + \mathcal{O}\bigl((-\mathrm{i}\eta)^{-1}\bigr),\quad\eta\to\infty.
\end{gather*}
Therefore, $\mathbf{F}^{(n)}_z(\eta,y,z)\mathbf{F}^{(n)}(\eta,y,z)^{-1}=\mathcal{O}\bigl((-\mathrm{i}\eta)^{-1}\bigr)$ as $\eta\to\infty$. Likewise, using~\eqref{eq:parametrix-at-0} we get
\begin{gather*}
\frac{\partial\breve{\mathbf{N}}^{(n),\mathrm{out}}}{\partial z}(\eta,y,z)\breve{\mathbf{N}}^{(n),\mathrm{out}}(\eta,y,z)^{-1}\\
\qquad {}=\bigg(\frac{\partial\mathbf{A}^{(n)}_0}{\partial z}(y,z) + \frac{\partial\mathbf{A}^{(n)}_1}{\partial z}(y,z) (-\mathrm{i}\eta) +\cdots\bigg)\\
\qquad\quad{}\times\bigl(\mathbf{A}^{(n)}_0(y,z)^{-1} - \mathbf{A}^{(n)}_0(y,z)^{-1}\mathbf{A}^{(n)}_1(y,z)\mathbf{A}_0^{(n)}(y,z)^{-1}(-\mathrm{i}\eta) + \cdots\bigr)=\mathcal{O}(1),
\end{gather*}
as $\eta\to 0$, and in the same limit
\begin{gather*}
(-\mathrm{i}\eta)^{-1/2}\breve{\mathbf{N}}^{(n),\mathrm{out}}(\eta,y,z)\sigma_3\breve{\mathbf{N}}^{(n),\mathrm{out}}(\eta,y,z)^{-1}\\
\qquad =\bigl(\mathbf{A}_0^{(n)}(y,z) + \mathbf{A}_1^{(n)}(y,z)(-\mathrm{i}\eta) + \cdots\bigr)\\
\qquad\quad{}\times (-\mathrm{i}\eta)^{-1/2}(-\mathrm{i}\eta)^{(-1)^n\sigma_3/4}\widetilde{\mathbf{E}}^{-1}\sigma_3\widetilde{\mathbf{E}}\cdot(-\mathrm{i}\eta)^{-(-1)^n\sigma_3/4}\\
\qquad\quad{}\times\bigl(\mathbf{A}^{(n)}_0(y,z)^{-1}-\mathbf{A}_0^{(n)}(y,z)^{-1}\mathbf{A}_1^{(n)}(y,z)\mathbf{A}_0^{(n)}(y,z)^{-1}(-\mathrm{i}\eta) + \cdots\bigr).
\end{gather*}
But, using the identity
\begin{equation*}
\widetilde{\mathbf{E}}^{-1}\sigma_3\widetilde{\mathbf{E}} = \begin{bmatrix}0 & \mathrm{e}^{-5\pi\mathrm{i}/3}\\\mathrm{e}^{5\pi\mathrm{i}/3} & 0\end{bmatrix},
\end{equation*}
the central factor becomes{\samepage
\begin{align}
& (-\mathrm{i}\eta)^{-1/2}(-\mathrm{i}\eta)^{(-1)^n\sigma_3/4}\widetilde{\mathbf{E}}^{-1}\sigma_3\widetilde{\mathbf{E}}\cdot(-\mathrm{i}\eta)^{-(-1)^n\sigma_3/4}
\nonumber\\
&\qquad =(-\mathrm{i}\eta)^{[(-1)^n\sigma_3-\mathbb{I}]/2}\begin{bmatrix}0 & \mathrm{e}^{-5\pi\mathrm{i}/3}\\\mathrm{e}^{5\pi\mathrm{i}/3} & 0\end{bmatrix} =\begin{cases}\mathbf{U} + \mathbf{L}\cdot(-\mathrm{i}\eta)^{-1},&\text{$n$ even}, \\\mathbf{U}\cdot(-\mathrm{i}\eta)^{-1} + \mathbf{L},&\text{$n$ odd},\end{cases} \label{eq:central-factor}
 \end{align}}%

 \pagebreak

\noindent
 where
 \begin{equation*}
 \mathbf{U}:=\begin{bmatrix}0&\mathrm{e}^{-5\pi\mathrm{i}/3}\\0&0\end{bmatrix}\qquad\text{and}\qquad\mathbf{L}:=\begin{bmatrix}0&0\\\mathrm{e}^{5\pi\mathrm{i}/3} & 0\end{bmatrix}.
 \end{equation*}
 Therefore, $\mathbf{F}^{(n)}_z(\eta,y,z)\mathbf{F}^{(n)}(\eta,y,z)^{-1}=\mathbf{T}\cdot(-\mathrm{i}\eta)^{-1} + \mathcal{O}(1)$ as $\eta\to 0$, wherein
 \begin{equation}
 \mathbf{T}=\mathbf{T}^{(n)}(y,z):=\begin{cases}\mathbf{A}^{(n)}_0(y,z)\mathbf{L}\mathbf{A}_0^{(n)}(y,z)^{-1},&\text{$n$ even,}\\\mathbf{A}_0^{(n)}(y,z)\mathbf{U}\mathbf{A}_0^{(n)}(y,z)^{-1},&\text{$n$ odd}.\end{cases}
 \label{eq:Z-define}
 \end{equation}
 It will also be convenient later to define the following related matrix
\begin{equation}
\widetilde{\mathbf{T}}=\widetilde{\mathbf{T}}^{(n)}(y,z):=\begin{cases}\mathbf{A}_0^{(n)}(y,z)\mathbf{U}\mathbf{A}_0^{(n)}(y,z)^{-1},&\text{$n$ even,}\\\mathbf{A}_0^{(n)}(y,z)\mathbf{L}\mathbf{A}_0^{(n)}(y,z)^{-1},&\text{$n$ odd}.\end{cases}
 \label{eq:Ztilde-define}
 \end{equation}
 These definitions imply that
 \begin{equation}
\mathbf{T}^2=\mathbf{0}\qquad\text{and}\qquad\widetilde{\mathbf{T}}^2=\mathbf{0}
\label{eq:ZsNilpotent}
\end{equation}
and that
\begin{equation}
\mathbf{T}\widetilde{\mathbf{T}}+\widetilde{\mathbf{T}}\mathbf{T}=\mathbb{I}.
\label{eq:anticommutator}
\end{equation}

 Note that, using~\eqref{eq:Un-approx} and $\det(\mathbf{A}_0)=1$, whether $n$ is even or odd one obtains the same formula for $\breve{U}_n$ in terms of $\mathbf{T}$:
 \begin{equation}
 \breve{U}_n=T_{11}=-T_{22}.
 \label{eq:breveU-Z}
 \end{equation}
 Comparing the expansions as $\eta\to\infty$ and as $\eta\to 0$ and multiplying on the right by $\mathbf{F}^{(n)}(\eta,y,z)$, we obtain the differential equation
 \begin{equation}
 \frac{\partial\mathbf{F}^{(n)}}{\partial z}(\eta,y,z)=\mathbf{T}\cdot(-\mathrm{i}\eta)^{-1}\mathbf{F}^{(n)}(\eta,y,z).
 \label{eq:dFdz}
 \end{equation}

\subsubsection{Implied differential equations for coefficients}
Combining~\eqref{eq:F-N} and~\eqref{eq:dFdz}, one obtains
 a differential equation for $\breve{\mathbf{N}}^{(n),\mathrm{out}}(\eta,y,z)$, namely
 \begin{equation}
 \frac{\partial\breve{\mathbf{N}}^{(n),\mathrm{out}}}{\partial z}(\eta,y,z)=\mathbf{T}\cdot (-\mathrm{i}\eta)^{-1}\breve{\mathbf{N}}^{(n),\mathrm{out}}(\eta,y,z) - (-\mathrm{i}\eta)^{-1/2}\breve{\mathbf{N}}^{(n),\mathrm{out}}(\eta,y,z)\sigma_3.
\label{eq:dNoutdz}
\end{equation}
Multiplying~\eqref{eq:dNoutdz} on the right by $\mathrm{e}^{ng(\eta,y)\sigma_3}\mathrm{e}^{-\mathrm{i} n\eta\sigma_3}\widetilde{\mathbf{E}}(-\mathrm{i}\eta)^{-(-1)^n\sigma_3/4}$ and using~\eqref{eq:parametrix-at-0} gives the convergent (by Remark~\ref{rem:zero}) series
\begin{gather*}
\sum_{j=0}^\infty \frac{\partial\mathbf{A}^{(n)}_j}{\partial z} (y,z)(-\mathrm{i}\eta)^j
 = \sum_{j=0}^\infty \mathbf{T}\mathbf{A}^{(n)}_j(y,z)(-\mathrm{i}\eta)^{j-1} \\
\qquad\quad{} - (-\mathrm{i}\eta)^{-1/2}\Bigg(\sum_{j=0}^{\infty}\mathbf{A}_j^{(n)}(y,z) (-\mathrm{i}\eta)^j\Bigg)(-\mathrm{i}\eta)^{(-1)^n\sigma_3/4}\widetilde{\mathbf{E}}^{-1}\sigma_3\widetilde{\mathbf{E}}\cdot(-\mathrm{i}\eta)^{-(-1)^n\sigma_3/4}\\
\qquad {} = \sum_{j=0}^\infty \mathbf{T}\mathbf{A}_j^{(n)}(y,z)(-\mathrm{i}\eta)^{j-1}-\sum_{j=0}^\infty\mathbf{A}_j^{(n)}(y,z)(-\mathrm{i}\eta)^j(-\mathrm{i}\eta)^{[(-1)^n\sigma_3-\mathbb{I}]/2}(\mathbf{U}+\mathbf{L}),
\end{gather*}
where we used~\eqref{eq:central-factor} in the second equality. Matching the coefficient of $(-\mathrm{i}\eta)^{-1}$ on both sides gives an identity according to the definition~\eqref{eq:Z-define} of $\mathbf{T}$. Matching the coefficient of $(-\mathrm{i}\eta)^0$ then gives
\begin{equation*}
\frac{\partial\mathbf{A}_0^{(n)}}{\partial z}(y,z)=\mathbf{T}\mathbf{A}_1^{(n)}(y,z)-\mathbf{A}_1^{(n)}(y,z)\mathbf{A}_0^{(n)}(y,z)^{-1}\mathbf{T}\mathbf{A}_0^{(n)}(y,z) - \widetilde{\mathbf{T}}\mathbf{A}_0^{(n)}(y,z).
\end{equation*}
Again using~\eqref{eq:Z-define} gives
\begin{gather}
\frac{\partial\mathbf{T}}{\partial z} = \frac{\partial\mathbf{A}_0^{(n)}}{\partial z}(y,z)\mathbf{A}_0^{(n)}(y,z)^{-1}\mathbf{T}-\mathbf{T}\frac{\partial\mathbf{A}_0^{(n)}}{\partial z}(y,z)\mathbf{A}_0^{(n)}(y,z)^{-1}\nonumber\\
\hphantom{\frac{\partial\mathbf{T}}{\partial z}}{} =\mathbf{T}\mathbf{A}_1^{(n)}(y,z)\mathbf{A}_0^{(n)}(y,z)^{-1}\mathbf{T}-\mathbf{A}_1^{(n)}(y,z)\mathbf{A}_0^{(n)}(y,z)^{-1}\mathbf{T}^2-\widetilde{\mathbf{T}}\mathbf{T} \nonumber\\
\hphantom{\frac{\partial\mathbf{T}}{\partial z}=}{} -\mathbf{T}^2\mathbf{A}_1^{(n)}(y,z)\mathbf{A}_0^{(n)}(y,z)^{-1}+\mathbf{T}\mathbf{A}_1^{(n)}(y,z)\mathbf{A}_0^{(n)}(y,z)^{-1}\mathbf{T}+\mathbf{T}\widetilde{\mathbf{T}}\nonumber\\
\hphantom{\frac{\partial\mathbf{T}}{\partial z}}{} =2\mathbf{T}\mathbf{A}_1^{(n)}(y,z)\mathbf{A}_0^{(n)}(y,z)^{-1}\mathbf{T}+\big[\mathbf{T},\widetilde{\mathbf{T}}\big].\label{eq:dZdz1}
\end{gather}
Here $[\mathbf{A},\mathbf{B}]:=\mathbf{AB}-\mathbf{BA}$ denotes the matrix commutator. Note that~\eqref{eq:dZdz1} is not a closed system of differential equations; in particular elements of $\mathbf{A}_1^{(n)}(y,z)$ appear on the right-hand side. To close the system, we next follow the suggestion of Its and investigate algebraic identities.

\subsection{Algebraic matrix identity}
Denoting $R(\eta,y):=-h_\eta(\eta,y,c_1(y))$, let
\begin{align}
\mathbf{G}^{(n)}(\eta,y,z):={} & R(\eta,y)\breve{\mathbf{N}}^{(n),\mathrm{out}}(\eta,y,z)\sigma_3\breve{\mathbf{N}}^{(n),\mathrm{out}}(\eta,y,z)^{-1} \nonumber\\
 = {} & R(\eta,y)\mathbf{F}^{(n)}(\eta,y,z)\sigma_3\mathbf{F}^{(n)}(\eta,y,z)^{-1},
\label{eq:G-def}
\end{align}
with $c=c_1(y)$ being the function defined in Lemma~\ref{lem:h}. Recalling~\eqref{eq:h-eta}, we have
\begin{align*}
R(\eta,y)^2 = f(\eta,y,c)=\frac{P(-\mathrm{i}\eta,y,c)}{(-\mathrm{i}\eta)^3} &= -1+(-\mathrm{i}\eta)^{-1}+c(-\mathrm{i}\eta)^{-2}-\tfrac{1}{4}y^2(-\mathrm{i}\eta)^{-3} \\ &= -\frac{(-\mathrm{i}\eta-s_1(y))(-\mathrm{i}\eta-s_2(y))(-\mathrm{i}\eta-s_3(y))}{(-\mathrm{i}\eta)^3}.
\end{align*}
Note that $\eta\mapsto R(\eta,y)$ changes sign across branch cuts where the jump matrix for $\breve{\mathbf{N}}^{(n),\mathrm{out}}(\eta,y,z)$ is off-diagonal, while $R(\eta,y)$ is analytic on $\Sigma_\infty^-$ and $\Sigma_0^+$ where the jump matrix is diagonal (this is the required special structure of the jump matrices for the method to apply). It follows easily that $\eta\mapsto\mathbf{G}^{(n)}(\eta,y,z)$ is analytic except at $\eta=0,\infty$.

Note also that
\begin{gather}
 R(\eta,y) =\begin{cases}
\mathrm{i} \Bigl(1-\tfrac{1}{2}(-\mathrm{i}\eta)^{-1} -\tfrac{1}{2}\bigl(c+\tfrac{1}{4}\bigr)(-\mathrm{i}\eta)^{-2} &\\
\qquad{} + \tfrac{1}{16}\bigl(2y^2-4c-1\bigr)(-\mathrm{i}\eta)^{-3} + \mathcal{O}\bigl((-\mathrm{i}\eta)^{-4}\bigr)\Bigr), &
\eta\to\infty, \\
-\tfrac{1}{2}\mathrm{i} y (-\mathrm{i}\eta)^{-3/2}\Bigl(1-\frac{2c}{y^2}(-\mathrm{i}\eta) -\frac{2(c^2+y^2)}{y^4}(-\mathrm{i}\eta)^2 &\\
\qquad{} +\frac{2(y^4-2c^3-2cy^2)}{y^6}(-\mathrm{i}\eta)^3+\mathcal{O}\bigl((-\mathrm{i}\eta)^{4}\bigr)\Bigr), &
\eta\to 0.
\end{cases}
\label{eq:R-expansions}
\end{gather}
From the fact that $\breve{\mathbf{N}}^{(n),\mathrm{out}}(\eta,y,z)\to\mathbb{I}$ as $\eta\to\infty$, analyticity of $\mathbf{G}^{(n)}(\eta,y,z)$ in $\mathbb{C}\setminus\{0\}$ implies that it has a Laurent expansion of the form
\begin{equation*}
\mathbf{G}^{(n)}(\eta,y,z)= \mathbf{G}^\infty_0 + \mathbf{G}^\infty_1\cdot (-\mathrm{i}\eta)^{-1} + \mathbf{G}^\infty_2\cdot(-\mathrm{i}\eta)^{-2} + \mathcal{O}\bigl((-\mathrm{i}\eta)^{-3}\bigr),\qquad\eta\to\infty.
\end{equation*}
Multiplying the definition~\eqref{eq:G-def} of $\mathbf{G}^{(n)}(\eta,y,z)$ on the right by $\breve{\mathbf{N}}^{(n)}(\eta,y,z)$, we insert the expansions~\eqref{eq:OuterParametrix-H-Expansion} and~\eqref{eq:R-expansions} and equate the coefficients of like powers of $(-\mathrm{i}\eta)^{-1}$ to obtain a~hierarchy of equations:
\begin{gather*}
\mathbf{G}^\infty_0=\mathrm{i}\sigma_3,
\\
\mathbf{G}^\infty_0\mathbf{C}_{1/2} = \mathrm{i}\mathbf{C}_{1/2}\sigma_3,
\\
\mathbf{G}^\infty_1+\mathbf{G}^\infty_0\mathbf{C}_1 = \mathrm{i}\mathbf{C}_1\sigma_3-\tfrac{1}{2}\mathrm{i}\sigma_3,
\\
\mathbf{G}^\infty_1\mathbf{C}_{1/2} +\mathbf{G}^\infty_0\mathbf{C}_{3/2}=\mathrm{i}\mathbf{C}_{3/2}\sigma_3-\tfrac{1}{2}\mathrm{i}\mathbf{C}_{1/2}\sigma_3,
\\
\mathbf{G}^\infty_2 + \mathbf{G}^\infty_1\mathbf{C}_1 + \mathbf{G}^\infty_0\mathbf{C}_2 = \mathrm{i}\mathbf{C}_2\sigma_3 -\tfrac{1}{2}\mathrm{i}\mathbf{C}_1\sigma_3-\tfrac{1}{2}\mathrm{i} \bigl(c+\tfrac{1}{4}\bigr)\sigma_3,
\end{gather*}
and so on. The first, third, and fifth equations give, in order,
\begin{gather*}
\mathbf{G}_0^\infty=\mathrm{i}\sigma_3,\\
\mathbf{G}_1^\infty=\mathrm{i}[\mathbf{C}_1,\sigma_3]-\tfrac{1}{2}\mathrm{i}\sigma_3,\\
\mathbf{G}_2^\infty=\mathrm{i}[\mathbf{C}_2,\sigma_3]-\tfrac{1}{2}\mathrm{i}[\mathbf{C}_1,\sigma_3]-\mathrm{i}[\mathbf{C}_1,\sigma_3]\mathbf{C}_1-\tfrac{1}{2}\mathrm{i}\bigl(c+\tfrac{1}{4}\bigr)\sigma_3.
\end{gather*}
According to~\eqref{eq:Cs-first} and \eqref{eq:C3/2-C1}, the second and fourth equations are trivial identities. Likewise, $\mathbf{G}^{(n)}(\eta,y,z)$ has a Laurent expansion about $\eta=0$ of the form
\begin{equation*}
\mathbf{G}^{(n)}(\eta,y,z)=\mathbf{G}^0_0\cdot (-\mathrm{i}\eta)^{-2} +\mathbf{G}^0_1\cdot (-\mathrm{i}\eta)^{-1} + \mathbf{G}^0_2 + \mathcal{O}((-\mathrm{i}\eta)),\qquad\eta\to 0.
\end{equation*}
Multiplying the definition of $\mathbf{G}^{(n)}(\eta,y,z)$ on the right by $\breve{\mathbf{N}}^{(n),\mathrm{out}}(\eta,y,z)\widetilde{\mathbf{E}}\cdot(-\mathrm{i}\eta)^{-(-1)^n\sigma_3/4}$ and using the expansions~\eqref{eq:parametrix-at-0} and~\eqref{eq:R-expansions} again gives a hierarchy of equations. To see them, first we expand the left-hand side:
\begin{gather*}
\mathbf{G}^{(n)}(\eta,y,z)\breve{\mathbf{N}}^{(n),\mathrm{out}}(\eta,y,z)\widetilde{\mathbf{E}}\cdot(-\mathrm{i}\eta)^{-(-1)^n\sigma_3/4}\\
\qquad{}=\bigl(\mathbf{G}^0_0\cdot (-\mathrm{i}\eta)^{-2} +\mathbf{G}^0_1\cdot (-\mathrm{i}\eta)^{-1} + \mathbf{G}^0_2 + \mathcal{O}((-\mathrm{i}\eta))\bigr)\\
\qquad\quad{}\times\bigl(\mathbf{A}_0^{(n)}(y,z) + \mathbf{A}_1^{(n)}(y,z)(-\mathrm{i}\eta) + \mathbf{A}_2^{(n)}(y,z)(-\mathrm{i}\eta)^2 + \mathcal{O}\bigl((-\mathrm{i}\eta)^3\bigr)\bigr)\\
\qquad{}=\mathbf{G}_0^0\mathbf{A}_0^{(n)}(y,z)(-\mathrm{i}\eta)^{-2} + \bigl(\mathbf{G}_1^0\mathbf{A}_0^{(n)}(y,z) + \mathbf{G}_0^0\mathbf{A}_1^{(n)}(y,z)\bigr)(-\mathrm{i}\eta)^{-1} \\
\qquad\quad{}+
\bigl(\mathbf{G}_2^0\mathbf{A}_0^{(n)}(y,z) + \mathbf{G}_1^0\mathbf{A}_1^{(n)}(y,z) + \mathbf{G}_0^0\mathbf{A}_2^{(n)}(y,z)\bigr) + \mathcal{O}((-\mathrm{i}\eta)),\qquad\eta\to 0.
\end{gather*}
Then we expand the right-hand side:
\begin{gather*}
R(\eta,y)\breve{\mathbf{N}}^{(n),\mathrm{out}}(\eta,y,z)\sigma_3\widetilde{\mathbf{E}}\cdot(-\mathrm{i}\eta)^{-(-1)^n\sigma_3/4}\\
\qquad {} =
R(\eta,y)\breve{\mathbf{N}}^{(n),\mathrm{out}}(\eta,y,z)\widetilde{\mathbf{E}}\cdot(-\mathrm{i}\eta)^{-(-1)^n\sigma_3/4}\\
 \qquad\quad{}\times
(-\mathrm{i}\eta)^{(-1)^n\sigma_3/4}\widetilde{\mathbf{E}}^{-1}\sigma_3\widetilde{\mathbf{E}}\cdot(-\mathrm{i}\eta)^{-(-1)^n\sigma_3/4}\\
\qquad{} =-\tfrac{1}{2}\mathrm{i} y \Big(1-\tfrac{2c}{y^2}(-\mathrm{i}\eta) -\tfrac{2(c^2+y^2)}{y^4}(-\mathrm{i}\eta)^2+\mathcal{O}\bigl((-\mathrm{i}\eta)^3\bigr)\Big)\\
 \qquad\quad{}\times\bigl(\mathbf{A}_0^{(n)}(y,z) + \mathbf{A}_1^{(n)}(y,z)(-\mathrm{i}\eta) + \mathbf{A}_2^{(n)}(y,z)(-\mathrm{i}\eta)^2 + \mathcal{O}\bigl((-\mathrm{i}\eta)^3\bigr)\bigr)\\
 \qquad\quad{}\times
(-\mathrm{i}\eta)^{[(-1)^n\sigma_3-3\mathbb{I}]/2}(\mathbf{U}+\mathbf{L}).
\end{gather*}
Recalling the definitions~\eqref{eq:Z-define} and \eqref{eq:Ztilde-define},
one sees that \[
(-\mathrm{i}\eta)^{[(-1)^n\sigma_3-3\mathbb{I}]/2}(\mathbf{U}+\mathbf{L})=\mathbf{A}_0^{(n)}(y,z)^{-1}\bigl(\mathbf{T}\cdot(-\mathrm{i}\eta)^{-2} +\widetilde{\mathbf{T}}\cdot(-\mathrm{i}\eta)^{-1}\bigr)\mathbf{A}_0^{(n)}(y,z).
\]
\pagebreak

\noindent
 This shows that
\begin{gather*}
R(\eta,y)\breve{\mathbf{N}}^{(n),\mathrm{out}}(\eta,y,z)\sigma_3\widetilde{\mathbf{E}}\cdot(-\mathrm{i}\eta)^{-(-1)^n\sigma_3/4}\\
\quad{} =-\frac{\mathrm{i} y}{2}\mathbf{T}\mathbf{A}_0^{(n)}(y,z) (-\mathrm{i}\eta)^{-2} \\
 \qquad{}+ \biggl(-\frac{\mathrm{i} y}{2} \mathbf{A}_1^{(n)}(y,z)\mathbf{A}_0^{(n)}(y,z)^{-1}\mathbf{T}\mathbf{A}_0^{(n)}(y,z)+\frac{\mathrm{i} c}{y}\mathbf{T}\mathbf{A}_0^{(n)}(y,z)-\frac{\mathrm{i} y}{2}\widetilde{\mathbf{T}}\mathbf{A}_0^{(n)}(y,z)\biggr)(-\mathrm{i}\eta)^{-1} \\
 \qquad{}-\frac{\mathrm{i} y}{2}\mathbf{A}_2^{(n)}(y,z)\mathbf{A}_0^{(n)}(y,z)^{-1}\mathbf{T}\mathbf{A}_0^{(n)}(y,z)+\frac{\mathrm{i} c}{y}\mathbf{A}_1^{(n)}(y,z)\mathbf{A}_0^{(n)}(y,z)^{-1}\mathbf{T}\mathbf{A}_0^{(n)}(y,z)\\
 \qquad{}+\frac{\mathrm{i} (c^2+y^2)}{y^4}\mathbf{T}\mathbf{A}_0^{(n)}(y,z) -\frac{\mathrm{i} y}{2}\mathbf{A}_1^{(n)}(y,z)\mathbf{A}_0^{(n)}(y,z)^{-1}\widetilde{\mathbf{T}}\mathbf{A}_0^{(n)}(y,z) +\frac{\mathrm{i} c}{y}\widetilde{\mathbf{T}}\mathbf{A}_0^{(n)}(y,z) \\
 \qquad{}+\mathcal{O}(-\mathrm{i}\eta),\qquad\eta\to 0.
\end{gather*}
Hence, matching the left and right-hand sides,
\begin{gather*}
\mathbf{G}_0^0=-\tfrac{1}{2}\mathrm{i} y\mathbf{T},
\\
\mathbf{G}_1^0=\tfrac{1}{2}\mathrm{i} y\big[\mathbf{T},\mathbf{A}_1^{(n)}(y,z)\mathbf{A}_0^{(n)}(y,z)^{-1}\big]+\frac{\mathrm{i} c}{y}\mathbf{T}-\tfrac{1}{2}\mathrm{i} y\widetilde{\mathbf{T}},
\end{gather*}
and so on. Since Liouville's theorem implies that $\mathbf{G}^{(n)}(\eta,y,z)$ is a Laurent polynomial in $-\mathrm{i}\eta$ of degree $(-2,0)$, we have the identities
\begin{equation*}
\mathbf{G}^\infty_2=\mathbf{G}_0^0,\qquad\mathbf{G}^\infty_1=\mathbf{G}_1^0,\qquad\mathbf{G}^\infty_0=\mathbf{G}_2^0.
\end{equation*}
When used with other facts obtained from~\eqref{eq:Z-define} and~\eqref{eq:Ztilde-define} such as~\eqref{eq:ZsNilpotent} and \eqref{eq:anticommutator}
these imply several interesting relations. For instance, $\mathbf{T}^2=\mathbf{0}$ implies that
\begin{equation*}
\mathbf{G}_1^0\mathbf{T}=\tfrac{1}{2}\mathrm{i} y\bigl(\mathbf{T}\mathbf{A}_1^{(n)}(y,z)\mathbf{A}_0^{(n)}(y,z)^{-1}\mathbf{T}-\widetilde{\mathbf{T}}\mathbf{T}\bigr)
\end{equation*}
and
\begin{equation*}
\mathbf{T}\mathbf{G}_1^0=\tfrac{1}{2}\mathrm{i} y\bigl(-\mathbf{T}\mathbf{A}_1^{(n)}(y,z)\mathbf{A}_0^{(n)}(y,z)^{-1}\mathbf{T}-\mathbf{T}\widetilde{\mathbf{T}}\bigr),
\end{equation*}
so using~\eqref{eq:anticommutator} the sum of these is $\mathbf{G}_1^0\mathbf{T}+\mathbf{T}\mathbf{G}_1^0=-\tfrac{1}{2}\mathrm{i} y\mathbb{I}$. However, by $\mathbf{G}_1^0=\mathbf{G}_1^\infty$, we get
\begin{equation}
\mathrm{i}\mathbf{T}\bigl([\mathbf{C}_1,\sigma_3]-\tfrac{1}{2}\sigma_3\bigr)+\mathrm{i}\bigl([\mathbf{C}_1,\sigma_3]-\tfrac{1}{2}\sigma_3\bigr)\mathbf{T}+\tfrac{1}{2}\mathrm{i} y\mathbb{I}=\mathbf{0}.
\label{eq:one-identity}
\end{equation}
The difference is $\big[\mathbf{G}_1^0,\mathbf{T}\big] = \tfrac{1}{2}\mathrm{i} y\bigl(2\mathbf{T}\mathbf{A}_1^{(n)}(y,z)\mathbf{A}_0^{(n)}(y,z)^{-1}\mathbf{T}+\big[\mathbf{T},\widetilde{\mathbf{T}}\big]\bigr)$. Again using $\mathbf{G}_1^0=\mathbf{G}_1^\infty$ gives
\begin{equation*}
\mathrm{i}\big[[\mathbf{C}_1,\sigma_3]-\tfrac{1}{2}\sigma_3,\mathbf{T}\big]=\tfrac{1}{2}\mathrm{i} y\bigl(2\mathbf{T}\mathbf{A}_1^{(n)}(y,z)\mathbf{A}_0^{(n)}(y,z)^{-1}\mathbf{T}+\big[\mathbf{T},\widetilde{\mathbf{T}}\big]\bigr).
\end{equation*}
This implies that the differential equation~\eqref{eq:dZdz1} can be written as a Lax commutator equation:
\begin{equation*}
\frac{\partial\mathbf{T}}{\partial z}=\frac{2}{y}\big[[\mathbf{C}_1,\sigma_3]-\tfrac{1}{2}\sigma_3,\mathbf{T}\big].
\label{eq:dZdz2}
\end{equation*}
In particular,
\begin{equation*}
\frac{\partial T_{11}}{\partial z}=-\frac{4}{y}(C_{1,12}T_{21}+T_{12}C_{1,21}),
\label{eq:dZ11dz}
\end{equation*}
which further implies that
\begin{align}
\bigg(\frac{\partial T_{11}}{\partial z}\bigg)^2 &= \frac{16}{y^2}\bigl(C_{1,12}^2T_{21}^2 + T_{12}^2C_{1,21}^2 + 2C_{1,12}C_{1,21}T_{12}T_{21}\bigr) \nonumber\\
&=\frac{16}{y^2}\bigl(C_{1,12}^2T_{21}^2 + T_{12}^2C_{1,21}^2-2C_{1,12}C_{1,21}T_{11}^2\bigr),\label{eq:FirstOrderSecondDegree1}
\end{align}
where on the second line we used $\det(\mathbf{T})=\mathrm{tr}(\mathbf{T})=0$.

\subsection{Scalar identity and completion of the proof of Theorem~\ref{thm:main-goal}}
The most remarkable identity stemming from the definition of $\mathbf{G}^{(n)}(\eta,y,z)$ comes from $\sigma_3^2=\mathbb{I}$ which implies that $\mathbf{G}^{(n)}(\eta,y,z)^2$ is the scalar (independent of both $n$ and $z$, and rational in $\eta$) $R(\eta,y)^2\mathbb{I}$. We can write $\mathbf{G}^{(n)}(\eta,y,z)$ in the form
\begin{align*}
\mathbf{G}^{(n)}(\eta,y,z)& =\mathbf{G}_0^0(-\mathrm{i}\eta)^{-2} + \mathbf{G}_1^\infty(-\mathrm{i}\eta)^{-1} +\mathbf{G}_0^\infty \\
& = -\tfrac{1}{2}\mathrm{i} y\mathbf{T}\cdot(-\mathrm{i}\eta)^{-2} +
\mathrm{i} \bigl([\mathbf{C}_1,\sigma_3]-\tfrac{1}{2}\sigma_3\bigr)(-\mathrm{i}\eta)^{-1} +\mathrm{i}\sigma_3.
\end{align*}
Therefore, using $\mathbf{T}^2=\mathbf{0}$,
\begin{gather*}
\mathbf{G}^{(n)}(\eta,y,z)^2 = \tfrac{1}{2}y\bigl(\mathbf{T}\bigl([\mathbf{C}_1,\sigma_3]-\tfrac{1}{2}\sigma_3\bigr)+\bigl([\mathbf{C}_1,\sigma_3]-\tfrac{1}{2}\sigma_3\bigr)\mathbf{T}\bigr)(-\mathrm{i}\eta)^{-3} \\
\hphantom{\mathbf{G}^{(n)}(\eta,y,z)^2 = }{}+ \bigl(\tfrac{1}{2}y(\mathbf{T}\sigma_3 + \sigma_3\mathbf{T})-\bigl([\mathbf{C}_1,\sigma_3]-\tfrac{1}{2}\sigma_3\bigr)^2\bigr)(-\mathrm{i}\eta)^{-2} \\
\hphantom{\mathbf{G}^{(n)}(\eta,y,z)^2 = }{}-\bigl(\bigl([\mathbf{C}_1,\sigma_3]-\tfrac{1}{2}\sigma_3\bigr)\sigma_3 + \sigma_3([\mathbf{C}_1,\sigma_3]-\tfrac{1}{2}\sigma_3)\bigr)(-\mathrm{i}\eta)^{-1} -\mathbb{I}.
\end{gather*}
Using~\eqref{eq:one-identity} and the fact that $[\mathbf{C}_1,\sigma_3]$ is off-diagonal, this becomes
\begin{gather*}
\mathbf{G}^{(n)}(\eta,y,z)^2 = -\tfrac{1}{4}y^2\mathbb{I}(-\mathrm{i}\eta)^{-3} + \bigl(\tfrac{1}{2}y(\mathbf{T}\sigma_3 + \sigma_3\mathbf{T})-\bigl([\mathbf{C}_1,\sigma_3]-\tfrac{1}{2}\sigma_3\bigr)^2\bigr)(-\mathrm{i}\eta)^{-2} \\
\hphantom{\mathbf{G}^{(n)}(\eta,y,z)^2 =}{}+\mathbb{I}(-\mathrm{i}\eta)^{-1}-\mathbb{I}.
\end{gather*}
Using $\mathrm{tr}(\mathbf{T})=0$, we verify that the coefficient of $(-\mathrm{i}\eta)^{-2}$ is also a multiple of the identity, and therefore
\begin{align*}
\mathbf{G}^{(n)}(\eta,y,z)^2 &{}= \bigl(-\tfrac{1}{4}y^2(-\mathrm{i}\eta)^{-3} + \bigl(yT_{11}+4C_{1,12}C_{1,21}-\tfrac{1}{4}\bigr)(-\mathrm{i}\eta)^{-2}+(-\mathrm{i}\eta)^{-1}-1\bigr)\mathbb{I} \\
&{}=R(\eta,y)^2\mathbb{I}.
\end{align*}
Then using this with $\eta=\mathrm{i} y/(4T_{11})$ and~\eqref{eq:FirstOrderSecondDegree1}, we have
\begin{gather*}
\bigg(\frac{\partial T_{11}}{\partial z}\bigg)^2+R\bigg(\frac{\mathrm{i} y}{4T_{11}},y\bigg)^2\\
\qquad {} =
\bigg(\frac{16}{y^2}\bigl(C_{1,12}^2T_{21}^2+T_{12}^2C_{1,21}^2\bigr)-\frac{32}{y^2}C_{1,12}C_{1,21}T_{11}^2\bigg) \\
\qquad\quad {}
 +\bigg(\frac{64}{y^2}C_{1,12}C_{1,21}T_{11}^2-\frac{4T_{11}^2}{y^2}+\frac{4T_{11}}{y}-1\bigg)\\
\qquad{} = \frac{16}{y^2}\bigl(C_{1,12}^2T_{21}^2+T_{12}^2C_{1,21}^2\bigr)+\frac{32}{y^2}C_{1,12}C_{1,21}T_{11}^2 -\frac{4T_{11}^2}{y^2}+\frac{4T_{11}}{y}-1\\
\qquad{} = \frac{16}{y^2}\bigl(C_{1,12}^2T_{21}^2+T_{12}^2C_{1,21}^2\bigr)-\frac{32}{y^2}C_{1,12}C_{1,21}T_{12}T_{21}
 -\frac{4T_{11}^2}{y^2}+\frac{4T_{11}}{y}-1\\
\qquad{} = \frac{16}{y^2}\bigl(C_{1,12}T_{21}-C_{1,21}T_{12}\bigr)^2-\frac{4T_{11}^2}{y^2}+\frac{4T_{11}}{y}-1,
\end{gather*}
where we used $\det(\mathbf{T})=\mathrm{tr}(\mathbf{T})=0$. But now, using the $(1,1)$-entry of the identity~\eqref{eq:one-identity} shows that
\begin{equation*}
\left(\frac{\partial T_{11}}{\partial z}\right)^2+R\left(\frac{\mathrm{i} y}{4T_{11}},y\right)^2=0.
\end{equation*}
In other words, recalling from~\eqref{eq:breveU-Z} that $T_{11}=\breve{U}=\breve{U}_n(z;y)$, we have shown that
\begin{equation}
\left(\frac{\partial\breve{U}}{\partial z}\right)^2=-R\left(\frac{\mathrm{i} y}{4\breve{U}},y\right)^2 = \frac{16}{y}\breve{U}^3-\frac{16c}{y^2}\breve{U}^2-\frac{4}{y}\breve{U}+1.
\label{eq:breveU-final}
\end{equation}
Comparing~\eqref{eq:breveU-first-order} and~\eqref{eq:breveU-final} shows that $\breve{U}(z)=\breve{U}_n(z;y)$ satisfies the expected differential equation equivalent to the Weierstra\ss\ equation~\eqref{eq:Weierstrass} with constant of integration $E$ connected to $y$ via~\eqref{eq:Ec},
which completes the proof of Theorem~\ref{thm:main-goal}.

\begin{Remark}
It is an interesting coincidence that the cubic polynomial in $\breve{U}$ appearing in the differential equation~\eqref{eq:breveU-final} is related to the rational function $R(\eta,y)^2=f(\eta,y,c_1(y))$ defining the underlying spectral curve (see~\eqref{eq:h-eta}) by $\eta=\mathrm{i} y/(4\breve{U})$. Similar correspondences have been noted with each application of this method; see~\cite{BuckinghamM14} for the original application to Painlev\'e-II, \cite{BothnerM20}~for an application to Painlev\'e-III (D$_6$), and~\cite{BuckinghamM-P4} for an application to Painlev\'e-IV.
\end{Remark}

\appendix

\section{Elementary illustration of the method}\label{app:toy-problem}

In this appendix we illustrate the method of proof of Theorem~\ref{thm:main-goal}
with a toy\footnote{In fact, Riemann--Hilbert Problem~\ref{toy-rhp}
arises in the description of unit-amplitude plane-wave solutions for the defocusing nonlinear Schr\"odinger equation $\mathrm{i} q_t +\frac{1}{2}q_{zz} -|q|^2q=0$ at time $t=0$ via $q(z,0)=2\mathrm{i} \lim_{\eta\to\infty}\eta\breve{N}_{12}(\eta,z)$. } example. Suppose the function $\breve{\mathbf{N}}(\eta,z)$ satisfies the
following Riemann--Hilbert problem:
\begin{rhp}[toy outer parametrix]\label{toy-rhp}
Given $z\in\mathbb{C}$, seek a $2\times 2$ matrix-valued function $(\mathbb{C}\setminus[-1,1])\ni\eta\mapsto\breve{\mathbf{N}}(\eta,z)$ with the following properties:
\begin{itemize}\itemsep=0pt
\item[]Analyticity: $\eta\mapsto\breve{\mathbf{N}}(\eta,z)$ is analytic in its domain of definition.
\item[]Jump condition: $\eta\mapsto\breve{\mathbf{N}}(\eta,z)$ takes continuous boundary values from the left $(+$ sub\-script$)$ and right $(-$ subscript$)$ on $(-1,1)$ oriented toward $\eta=1$ except at the endpoints $\eta=\pm 1$. The jump condition relating the boundary values is
\[
\breve{\mathbf{N}}_+(\eta,z)=\breve{\mathbf{N}}_-(\eta,z)\begin{bmatrix}0 & \mathrm{e}^{2\mathrm{i} z\eta}\\
-\mathrm{e}^{-2\mathrm{i} z\eta} & 0\end{bmatrix},\qquad z\in (-1,1).
\]
\item[]Normalization: $\breve{\mathbf{N}}(\eta,z)\to\mathbb{I}$ as $\eta\to\infty$.
\item[]Endpoint behavior: $\eta\mapsto\breve{\mathbf{N}}(\eta,z)$ is allowed to blow up like a negative one-fourth power near each endpoint $\eta=\pm 1$.
\end{itemize}
\end{rhp}

It is easy to check that this problem has a unique solution for every $z\in\mathbb{C}$ given explicitly by%
\begin{equation}
\breve{\mathbf{N}}(\eta,z)=\frac{1}{\sqrt{2}}\begin{bmatrix}1 & \mathrm{i}\\\mathrm{i} & 1\end{bmatrix}\left(\frac{\eta-1}{\eta+1}\right)^{\sigma_3/4}\left(\frac{1}{\sqrt{2}}\begin{bmatrix}1&\mathrm{i}\\\mathrm{i} & 1\end{bmatrix}\right)^{-1}
\mathrm{e}^{-\mathrm{i} z (\eta-R(\eta))\sigma_3},
\label{eq:Toy-Exact}
\end{equation}
where the diagonal matrix power is defined as the principal branch, $R(\eta)^2=\eta^2-1$, and $R(\eta)$ is analytic for $\eta\not\in [-1,1]$ with $R(\eta)=\eta-\frac{1}{2}\eta^{-1}+\mathcal{O}\bigl(\eta^{-3}\bigr)$ as $\eta\to\infty$.
If we let $\breve{\mathbf{N}}^{(k)}(z)$ denote the coefficients in the convergent Laurent expansion
\begin{equation}
\breve{\mathbf{N}}(\eta,z)=\mathbb{I}+\sum_{k=1}^\infty\breve{\mathbf{N}}^{(k)}(z)\eta^{-k},\qquad |\eta|>1,
\label{eq:Toy-expansion}
\end{equation}
then it is straightforward to obtain from~\eqref{eq:Toy-Exact} that
\[
\breve{N}^{(1)}_{11}(z)=-\frac{1}{2}\mathrm{i} z,\qquad \breve{N}^{(1)}_{22}(z)=\frac{1}{2}\mathrm{i} z.
\]
In particular, this implies that the diagonal elements of $\breve{\mathbf{N}}^{(1)}(z)$ satisfy simple differential equations:
\begin{equation}
\frac{\mathrm{d}\breve{N}^{(1)}_{11}}{\mathrm{d} z}=-\frac{1}{2}\mathrm{i},\qquad\frac{\mathrm{d}\breve{N}^{(1)}_{22}}{\mathrm{d} z}=\frac{1}{2}\mathrm{i}.
\label{eq:diagonal-ODE-Toy}
\end{equation}

We will now derive the differential equations~\eqref{eq:diagonal-ODE-Toy} without
using the explicit solution formula~\eqref{eq:Toy-Exact}.
(Analogously, in the proof of Theorem~\ref{thm:main-goal}, we directly derive
\eqref{eq:breveU-first-order}.)
First set $\mathbf{F}(\eta,z):=\breve{\mathbf{N}}(\eta,z)\mathrm{e}^{\mathrm{i}
z\eta\sigma_3}$, and observe that
\[
\mathbf{Z}(\eta,z):=\frac{\partial\mathbf{F}}{\partial z}(\eta,z)\mathbf{F}(\eta,z)^{-1}=\frac{\partial \breve{\mathbf{N}}}{\partial z}(\eta,z)\breve{\mathbf{N}}(\eta,z)^{-1}+\mathrm{i} \eta\breve{\mathbf{N}}(\eta,z)\sigma_3\breve{\mathbf{N}}(\eta,z)^{-1}
\]
is an entire function. Since $\breve{\mathbf{N}}_z(\eta,z)=\mathcal{O}\bigl(\eta^{-1}\bigr)$ as $\eta\to\infty$ because $\breve{\mathbf{N}}(\eta,z)$ itself tends to a~constant matrix $\mathbb{I}$ as $\eta\to\infty$, we get the expansion
\begin{align*}
\mathbf{Z}(\eta,z) =\mathrm{i} \eta\breve{\mathbf{N}}(\eta,z)\sigma_3\breve{\mathbf{N}}(\eta,z)^{-1}+\mathcal{O}\bigl(\eta^{-1}\bigr)
=\mathrm{i} \eta\sigma_3 +\mathrm{i} \big[\breve{\mathbf{N}}^{(1)}(z),\sigma_3\big] + \mathcal{O}\bigl(\eta^{-1}\bigr)
\end{align*}
as $\eta\to\infty$, so by Liouville's theorem,
\[
\mathbf{Z}(\eta,z)=\mathrm{i} \eta\sigma_3 +\mathrm{i} \big[\breve{\mathbf{N}}^{(1)}(z),\sigma_3\big].
\]
Therefore, $\mathbf{F}(\eta,z)$ satisfies the differential equation
\[
\frac{\partial\mathbf{F}}{\partial z}(\eta,z)=\mathbf{Z}(\eta,z)\mathbf{F}(\eta,z)
\]
or, in terms of $\breve{\mathbf{N}}(\eta,z)$ itself,
\begin{equation}
\frac{\partial\breve{\mathbf{N}}}{\partial z}(\eta,z)=\mathrm{i} \big[\breve{\mathbf{N}}^{(1)}(z),\sigma_3\big]\breve{\mathbf{N}}(\eta,z) -\mathrm{i}\eta\big[\breve{\mathbf{N}}(\eta,z),\sigma_3\big].
\label{eq:Toy-eta-ODE}
\end{equation}
Using the expansion~\eqref{eq:Toy-expansion} in~\eqref{eq:Toy-eta-ODE},
the diagonal terms proportional to $\eta^{-1}$ give
\begin{equation}
\frac{\mathrm{d}\breve{N}^{(1)}_{11}}{\mathrm{d} z}(z) = -2\mathrm{i}\breve{N}^{(1)}_{12}(z)\breve{N}^{(1)}_{21}(z),\qquad
\frac{\mathrm{d}\breve{N}^{(1)}_{22}}{\mathrm{d} z}(z) = 2\mathrm{i}\breve{N}^{(1)}_{12}(z)\breve{N}^{(1)}_{21}(z).
\label{eq:diagonal-ODE-pre-Toy}
\end{equation}
To close the system, we define a matrix function $\mathbf{G}(\eta,z)$ by
\begin{equation}
\mathbf{G}(\eta,z):=R(\eta)\breve{\mathbf{N}}(\eta,z)\sigma_3\breve{\mathbf{N}}(\eta,z)^{-1}.
\label{eq:Toy-G-def}
\end{equation}
Using the jump condition satisfied by $\eta\mapsto\breve{\mathbf{N}}(\eta,z)$, one checks easily that $\mathbf{G}_+(\eta,z)=\mathbf{G}_-(\eta,z)$ holds for $z\in (-1,1)$, and from the behavior of $\eta\mapsto\breve{\mathbf{N}}(\eta,z)$ at the endpoints of $[-1,1]$ one sees that $\mathbf{G}(\eta,z)$ is bounded at $z=\pm 1$. It follows that for each $z\in\mathbb{C}$, $\eta\mapsto \mathbf{G}(\eta,z)$ is an entire function. The expansion~\eqref{eq:Toy-expansion} then implies corresponding asymptotic behavior of $\mathbf{G}(\eta,z)$:
\[
\mathbf{G}(\eta,z)=\eta\sigma_3 + \big[\breve{\mathbf{N}}^{(1)}(z),\sigma_3\big] + \mathcal{O}\bigl(\eta^{-1}\bigr),\qquad\eta\to\infty
\]
and hence by Liouville's theorem we obtain the exact identity
\begin{equation}
\mathbf{G}(\eta,z)=\eta\sigma_3 + \big[\breve{\mathbf{N}}^{(1)}(z),\sigma_3\big].
\label{eq:G-identity-Toy}
\end{equation}
Moreover, since $\sigma_3^2=\mathbb{I}$ and $R(\eta)^2=z^2-1$, it follows from~\eqref{eq:Toy-G-def} that
\[
\mathbf{G}(\eta,z)^2=R(\eta)^2\mathbb{I} = \eta^2\mathbb{I}-\mathbb{I},
\]
while on the other hand, by squaring the identity~\eqref{eq:G-identity-Toy} and using the fact that for any $2\times 2$ matrix $\mathbf{M}$ one has $\sigma_3[\mathbf{M},\sigma_3]+[\mathbf{M},\sigma_3]\sigma_3=\mathbf{0}$,
\[
\mathbf{G}(\eta,z)^2 = \eta^2\mathbb{I} + \big[\breve{\mathbf{N}}^{(1)}(z),\sigma_3\big]^2.
\]
Comparing these two representations of $\mathbf{G}(\eta,z)^2$, we therefore obtain
\[
\big[\breve{\mathbf{N}}^{(1)}(z),\sigma_3\big]^2=-\mathbb{I}\implies 4\breve{N}^{(1)}_{12}(z)\breve{N}^{(1)}_{21}(z)=1.
\]
Using this identity in~\eqref{eq:diagonal-ODE-pre-Toy} closes the system and yields~\eqref{eq:diagonal-ODE-Toy}.

\section{Proof of Lemma~\ref{lem:h}}

\begin{proof}
Suppose that the roots $\eta=\mathrm{i} s_j$, $j=1,2,3$ are distinct, and that the Boutroux conditions~\eqref{eq:Boutroux} hold (this will be justified later via a continuation argument). With an integration constant selected so that $\operatorname{Re}(h(\eta,y,c))=0$ at any one of the roots, the zero level set $K$ of $\operatorname{Re}(h(\eta,y,c))$ is well defined and it contains the closure $K'$ of the union of \emph{critical trajectories} emanating from the points $\eta=\mathrm{i} s_j$, $j=1,2,3$, which are the curves along which $f(\eta,y,c)\,\mathrm{d}\eta^2<0$, where $f$ is defined by~\eqref{eq:h-eta}. We claim that $K'$ has the following properties.
\begin{itemize}\itemsep=0pt
\item $K'$ is a connected set consisting of six simple arcs pairwise disjoint except for their endpoints:
\begin{itemize}\itemsep=0pt
\item One arc joining the origin $\eta=0$ to one of the three points that we label as $\eta=\mathrm{i} s_1$. We take this arc to be the branch cut $\Sigma_{0,1}$.
\item One arc joining the other two points, $\eta=\mathrm{i} s_2$ and $\eta=\mathrm{i} s_3$. We take this arc to be the branch cut $\Sigma_{2,3}$.
\item Two arcs joining $\eta=\mathrm{i} s_1$ either to the same point that we label as $\eta=\mathrm{i} s_2$ (case (i)) or one each to $\eta=\mathrm{i} s_2$ and $\eta=\mathrm{i} s_3$ (case (ii)).
\item Two unbounded arcs tending to $\eta=\infty$ parallel to the real line, one in the right half-plane and one in the left half-plane.
\end{itemize}
\item In case (i), the region bounded by the two arcs joining $\eta=\mathrm{i} s_1$ with $\eta=\mathrm{i} s_2$ contains the origin $\eta=0$ and both unbounded arcs emanate from $\eta=\mathrm{i} s_3$. In case (ii), the region bounded by the arc joining $\eta=\mathrm{i} s_1$ with $\eta=\mathrm{i} s_2$, the arc joining $\eta=\mathrm{i} s_1$ with $\eta=\mathrm{i} s_3$, and the arc $\Sigma_{2,3}$ contains the origin $\eta=0$ and one unbounded arc emanates from each of $\eta=\mathrm{i} s_2$ and $\eta=\mathrm{i} s_3$.
\end{itemize}
Locally, near each of the simple roots $\eta=\mathrm{i} s_j$, $j=1,2,3$, of $\eta\mapsto f(\eta,y,c)$, $K'$ consists of a union of three trajectories emanating from $\eta=\mathrm{i} s_j$ in directions separated by equal angles of $2\pi/3$. Given an index $j=1,2,3$, each of the three trajectories emanating from $\eta=\mathrm{i} s_j$ terminates in the other direction at $\eta=\mathrm{i} s_1$, $\eta=\mathrm{i} s_2$, $\eta=\mathrm{i} s_3$, $\eta=0$, or is unbounded in which case it tends to $\eta=\infty$ asymptotically horizontally. This is because otherwise the trajectory would be \emph{divergent} and hence \emph{recurrent}~\cite[Theorem 11.1]{Strebel84}. But the closure of a~recurrent trajectory contains a~nonempty domain in $\mathbb{C}$ and since $\operatorname{Re}(h(\eta,y,c))=0$ on the trajectory, this harmonic function would vanish identically on $\mathcal{R}$ (the Riemann surface of $h_\eta(\eta,y,c)$, i.e., the spectral curve), which is a contradiction because $h_\eta(\eta,y,c)$ is not identically zero. Taking into account the Boutroux conditions~\eqref{eq:Boutroux} which imply that $K'\subset K$, similar local analysis shows that there can be at most one critical trajectory terminating at $\eta=0$ and at most one unbounded critical trajectory tending horizontally to $\eta=\infty$ in each of the left and right half-planes.

To work out the global trajectory structure in order to prove the claim, it is easiest to first assume that $y>0$ and $c\in\mathbb{R}$, in which case it is easy to see that $h_\eta(-\eta^*,y,c)=h_\eta(\eta,y,c)^*$ provided that $\Sigma_{0,1}$ and $\Sigma_{2,3}$ are taken to be symmetric in the imaginary $\eta$-axis, which we will also assume. Moreover we either have (making a choice of labeling of the roots of $P$) $s_1<0<s_2<s_3$ or $s_1<0$ and $s_3=s_2^*$. In either configuration the condition $I_{2,3}=0$ holds automatically, and $c\in\mathbb{R}$ is presumed to be determined from the remaining real condition $I_{1,2}=0$. We examine the two configurations in turn.

If $s_1<0<s_2<s_3$, since $f(\eta,y,c)=h_\eta(\eta,y,c)^2>0$ holds for $\eta\in\mathrm{i}\mathbb{R}$ between $\eta=\mathrm{i} s_1$ and $\eta=0$ as well as between $\eta=\mathrm{i} s_2$ and $\eta=\mathrm{i} s_3$, these intervals of the imaginary axis are critical trajectories. Since elsewhere on the imaginary axis we have ${f(\eta,y,c)=h_\eta(\eta,y,c)^2<0}$, $\operatorname{Re}(h(\eta,y,c))$ is strictly monotone as $\eta$ varies in these intervals of $\mathrm{i}\mathbb{R}$. Therefore, in this configuration there are no points of either $K$ or $K'\subset K$ on the imaginary axis outside the two critical trajectories. Since $K'$ is symmetric in reflection through the imaginary axis, and since exactly one critical trajectory goes to $\eta=\infty$ in each half-plane, there are only three possibilities:
\begin{itemize}\itemsep=0pt
\item The two remaining trajectories emanating from $\eta=\mathrm{i} s_1$ tend to infinity in opposite half-planes, and there is a symmetric pair of arcs in each half-plane joining the points $\eta=\mathrm{i} s_2$ and $\eta=\mathrm{i} s_3$. However, since $\operatorname{Re}(h(\eta,y,c))$ is harmonic between the imaginary axis and each of these arcs and vanishes on each critical trajectory, this would imply by the maximum principle that $\operatorname{Re}(h(\eta,y,c))\equiv 0$ in each of these domains. This is a contradiction since $h_\eta(\eta,y,c)$ does not vanish identically.
\item The two remaining trajectories emanating from $\eta=\mathrm{i} s_2$ tend to infinity in opposite half-planes, and there is a symmetric pair of arcs in each half-plane joining the points $\eta=\mathrm{i} s_1$ and $\eta=\mathrm{i} s_3$. However, this would imply a crossing of two different trajectories at a point in each half-plane where $f(\eta,y,c)$ is finite and nonzero, which cannot occur.
\item Therefore, the remaining possibility must hold, namely that the two remaining trajectories emanating from $\eta=\mathrm{i} s_3$ tend to infinity in opposite half-planes, and there is a symmetric pair of arcs in each half-plane joining the points $\eta=\mathrm{i} s_1$ and $\eta=\mathrm{i} s_2$.
\end{itemize}
This shows that the claimed structure holds in case (i) when $s_1<0<s_2<s_3$.

If instead $s_1<0$ and $s_3=s_2^*$, since $f(\eta,y,c)=h_\eta(\eta,y,c)^2>0$ holds for $\eta\in \mathrm{i} \mathbb{R}$ between $\eta=\mathrm{i} s_1$ and $\eta=0$, this interval of the imaginary axis is a critical trajectory, while in the intervals between $\eta=-\mathrm{i}\infty$ and $\eta=\mathrm{i} s_1$ and between $\eta=0$ and $\eta=+\mathrm{i}\infty$ we have $f(\eta,y,c)=h_\eta(\eta,y,c)^2<0$ so $\operatorname{Re}(h(\eta,y,c))$ is strictly monotone. This implies that there are no points of either $K$ or $K'\subset K$ on the imaginary axis below $\eta=\mathrm{i} s_1$, but because $\operatorname{Re}(h(\eta,y,c))$ necessarily changes sign on the positive imaginary axis due to the singularity at the origin and the linear growth at infinity there is exactly one point of $K$ there, which may belong to $K'$. In fact, this point does indeed belong to $K'$, because otherwise at least two of the critical trajectories emanating from each of the points $\eta=\mathrm{i} s_2$ and $\eta=\mathrm{i} s_3=-(\mathrm{i} s_2)^*$ must tend to infinity in the half-plane containing the point because only one of them can terminate at $\eta=\mathrm{i} s_1$; this contradicts the fact that exactly one critical trajectory tends to infinity in each half-plane. So the distinguished point in the imaginary interval between $\eta=0$ and $\eta=+\mathrm{i}\infty$ belongs to $K'$ and lies on a critical trajectory crossing the imaginary axis horizontally and connecting $\eta=\mathrm{i} s_2$ and $\eta=\mathrm{i} s_3=-(\mathrm{i} s_2)^*$. The remaining two trajectories emanating from each of these points necessarily tend to $\eta=\mathrm{i} s_1$ and $\eta=\infty$ without crossing. This shows that the claimed structure of $K'$ holds in case (ii) when $s_1<0$ and $s_3=s_2^*$.

We next show that whenever $0<y<y_\mathrm{c}\approx 0.29177$, where $y_\mathrm{c}$ is the critical value defined in~\cite[Section 4.6]{BuckinghamM22}, there exists a unique $c=c_1(y)\in\mathbb{R}$ for which the conditions~\eqref{eq:Boutroux} (really just $I_{1,2}=0$ as $I_{2,3}=0$ is automatic) hold with root configuration $s_1<0<s_2<s_3$ and hence $K'$ has the claimed structure in case (i). To do this, we first suppose that $y>0$ and choose $c\in\mathbb{R}$ differently, so that the cubic $\mu\mapsto P(\mu,y,c)$ defined in~\eqref{eq:h-eta} has a simple root $\mu=s$ and a double root $\mu=d$.
Then by setting $P(\mu,y,c)=-(\mu-d)^2(\mu-s)$ one sees that $d=(1-s)/2$ and that $s$ satisfies the cubic equation $s(s-1)^2=-y^2$, while $c=-d^2-2ds=\frac{1}{4}(3s^2-2s-1)$. The condition $y>0$ implies that the equation $s(s-1)^2=-y^2$ has one real and two complex-conjugate solutions for $s$. But if $s=u+\mathrm{i} v$ with $v\neq 0$, then $\operatorname{Im}(c)=\frac{1}{2}(3u-1)v$ which vanishes for $v\neq 0$ only if $u=\frac{1}{3}$. Then $\operatorname{Im}\bigl(y^2\bigr)=-\operatorname{Im}\bigl(s(s-1)^2\bigr)=v^3\neq 0$ which contradicts $y>0$. Therefore, the conditions $y>0$ and $c\in\mathbb{R}$ require that we select the real root $s=s(y)<0$ of $s(s-1)^2=-y^2$ and then $d=d(y)>\frac{1}{2}$ is also real and $c=c_0(y)$ is a corresponding well-defined real number. In this double-root configuration, the function $\eta\mapsto h_\eta(\eta,y,c_0(y))$ is analytic except on the imaginary segment between $\eta=\mathrm{i} s(y)$ and $\eta=0$. Choosing an integration constant so that $\operatorname{Re}(h(\eta,y,c_0(y)))=0$ for $\eta=\mathrm{i} s(y)$, the function $\operatorname{Re}(h(\eta,y,c_0(y)))$ is well defined by contour integration and it is harmonic except on the branch cut for $h_\eta(\eta,y,c_0(y))$. According to~\cite[Section 4.6]{BuckinghamM22}, if $y>0$ and $y\neq y_\mathrm{c}$, then $y-y_\mathrm{c}$ and $\operatorname{Re}(h(\eta,y,c_0(y)))$ always have the same sign for $\eta=\mathrm{i} d(y)$. Now $c=c_0(y)$ is necessarily a real root of the cubic discriminant of $\mu\mapsto P(\mu,y,c)$, which is proportional by a positive numerical factor to $64 c^3+16c^2+72y^2c + 16y^2-27y^4$. The discriminant of this latter polynomial with respect to $c$ is proportional by a positive numerical factor to $-y^2\bigl(4+27y^2\bigr)^3$ which is strictly negative for $y>0$, hence $c=c_0(y)$ is a simple root of the cubic discriminant and it is the only real root thereof. Clearly the cubic discriminant of $\mu\mapsto P(\mu,y,c)$ has the same sign as $c$ when $y>0$ is fixed and $c\in\mathbb{R}$ is large. Hence $\mu\mapsto P(\mu,y,c)$ has three real distinct roots whenever $c>c_0(y)$ and has only one real root (simple) whenever~${c<c_0(y)}$.

Now returning to the general case of $y>0$ and $c\in\mathbb{R}$ arbitrary (so that $P(\mu,y,c)$ need not have a double root), we wish to solve the equation $I_{1,2}=0$ for $c$ using the above information. A~simple calculation shows that $\frac{\partial}{\partial c}h_\eta(\eta,y,c)^2 = -\eta^{-2}$, which implies that when $c\in\mathbb{R}$ and~${y>0}$, \looseness=1
\begin{equation*}
\frac{\partial I_{1,2}}{\partial c}=\operatorname{Re}\bigg(\int_{\mathrm{i} s_1}^{\mathrm{i} s_2}\frac{\partial^2 h}{\partial\eta\partial c}(\eta,y,c)\,\mathrm{d} y\bigg)=-\frac{1}{2}\operatorname{Re}\bigg(\int_{\mathrm{i} s_1}^{\mathrm{i} s_2}\frac{\mathrm{d}\eta}{\eta^2 h_\eta(\eta,y,c)}\bigg),
\end{equation*}
because $h_\eta(\eta,y,c)$ vanishes at $\eta=\mathrm{i} s_1$ and $\eta=\mathrm{i} s_2$. Some contour deformations show that regardless of whether $c<c_0(y)$ or $c>c_0(y)$, this derivative can be written in a universal form:
\begin{equation*}
\frac{\partial I_{1,2}}{\partial c}=\frac{1}{4}\int_\mathbb{R}\frac{\chi_{P>0}(\mu)\,\mathrm{d}\mu}{\sqrt{P(\mu,y,c)}},
\end{equation*}
where $\chi_{P>0}(\mu)$ is the characteristic function of the union of intervals of $\mathbb{R}$ on which $\mu\mapsto P(\mu,y,c)$ is positive, and the square root is positive. Therefore, in both cases $c<c_0(y)$ and $c>c_0(y)$ the partial derivative of $I_{1,2}$ with respect to $c\in\mathbb{R}$ for $y>0$ is positive (the difference is that the support of $\chi_{P>0}(\mu)$ consists of two intervals $(-\infty,s_1)\cup (0,+\infty)$ for $c<c_0(y)$ and of three intervals $(-\infty,s_1)\cup (0,s_2)\cup (s_3,+\infty)$ for $c>c_0(y)$). Now, the real-valued function $c\mapsto I(c):=I_{1,2}$ is certainly continuous as a function of $c\in\mathbb{R}$ and is continuously differentiable for $c\neq c_0(y)$ with positive derivative. We also know that $I(c_0(y))=\operatorname{Re}(h(\mathrm{i} d(y),y,c_0(y)))<0$ for $0<y<y_\mathrm{c}$. However, it is also true that eventually $I(c)>0$ as $c\to+\infty$. To see this, first note that the roots of $P(\mu,y,c)$ for large $c>0$ with fixed $y>0$ are $s_1=-c^{1/2}+O(1)$, $s_2=\frac{1}{4}y^2c^{-1}+O\bigl(c^{-2}\bigr)$, and $s_3=c^{1/2}+O(1)$. Then one rescales the integrand of $I_{1,2}$ by $\eta=s_2w$ and notices that by contour deformations, the leading term of $I_{1,2}$ as $c\to+\infty$ with $y>0$ fixed is computed as a residue. Indeed, letting $C_\mathrm{L}$ (resp.\ $C_\mathrm{R}$) denote a contour in the left (resp.\ right) half-plane beginning at a point on the imaginary axis between $\eta=\mathrm{i} s_1$ and $\eta=0$ and terminating at a point on the imaginary axis between $\eta=\mathrm{i} s_2$ and $\eta=\mathrm{i} s_3$, letting $L$ denote a~positively-oriented loop surrounding the imaginary interval between $w=0$ and $w=\mathrm{i}$, and using principal branches for all power functions,
\begin{align*}
I_{1,2}&=\operatorname{Re}\left(\int_{\mathrm{i} s_1}^{\mathrm{i} s_2}h_\eta(\eta,y,c)\,\mathrm{d}\eta\right) \\ &=
\frac{1}{2}\operatorname{Re}\left(\int_{C_\mathrm{L}\cup C_\mathrm{R}}\left[-\mathrm{i}(-\mathrm{i}\eta-s_3)^{1/2}(-\mathrm{i}\eta-s_2)^{1/2}(-\mathrm{i}\eta)^{-3/2}(-\mathrm{i}\eta-s_1)^{1/2}\right]\,\mathrm{d}\eta\right)\\
&=\frac{1}{2}\operatorname{Re}\left(\oint_{(-C_\mathrm{L})\cup C_\mathrm{R}}\left[-\mathrm{i} (-\mathrm{i}\eta-s_1)^{1/2}\left(\frac{\eta-\mathrm{i} s_2}{\eta}\right)^{1/2}(\mathrm{i}\eta+s_3)^{1/2}\right]\frac{\mathrm{d}\eta}{\eta}\right)\\
&=\frac{1}{2}\sqrt{c}\operatorname{Re}\left(\oint_{L}\left[-\mathrm{i}\left(-\mathrm{i}\frac{s_2}{\sqrt{c}}w-\frac{s_1}{\sqrt{c}}\right)^{1/2}\left(\frac{w-\mathrm{i}}{w}\right)^{1/2}\left(\mathrm{i}\frac{s_2}{\sqrt{c}}+\frac{s_3}{\sqrt{c}}\right)^{1/2}\right]\,\frac{\mathrm{d} w}{w}\right)\\
&=\frac{1}{2}\sqrt{c}\left[\operatorname{Re}\left(\oint_L\left[-\mathrm{i}\left(\frac{w-\mathrm{i}}{w}\right)^{1/2}\right]\frac{\mathrm{d} w}{w}\right)+o(1)\right] \\ 
&=\frac{1}{2}\sqrt{c}\left[\operatorname{Re}\left(2\pi\mathrm{i}\mathop{\mathrm{Res}}_{w=\infty}
\left[-\mathrm{i}\left(\frac{w-\mathrm{i}}{w}\right)^{1/2}\frac{1}{w}\right]\right)+o(1)\right] \\ 
&=\pi\sqrt{c}+o(\sqrt{c}),\qquad c\to+\infty.
\end{align*}
Therefore if $0<y<y_\mathrm{c}$, $I(c)<0$ for $c=c_0(y)$ while $I(c)>0$ in the limit $c\to+\infty$, and $c\mapsto I(c)$ is strictly increasing for $c>c_0(y)$. It follows from the intermediate value theorem that there is a unique solution $c=c_1(y)>c_0(y)$ of $I(c)=I_{1,2}=0$, and the inequality $c_1(y)>c_0(y)$ implies that the roots of $\mu\mapsto P(\mu,y,c)$ are all real and hence $K'$ has the claimed structure in case (i) whenever $0<y<y_\mathrm{c}$.

Now we continue this solution into the complex $y$-plane. Since for $0<y<y_\mathrm{c}$ the roots of $\mu\mapsto P(\mu,y,c_1(y))$ are distinct, the solution $c=c_1(y)$ of the system~\eqref{eq:Boutroux} can be continued uniquely by the implicit function theorem to some maximal domain of the complex $y$-plane containing the real interval $(0,y_\mathrm{c})$ (however note that $c_1(y)$ is not an analytic function of $y$ in this domain). The boundary of this domain consists of all points $y$ for which $c_1(y)=c_0(y)$, i.e., under continuation of the solution of the Boutroux conditions~\eqref{eq:Boutroux} one arrives at a degenerate spectral curve with $\mu\mapsto P(\mu,y,c)$ having a double root. This boundary was obtained in~\cite[Section 4.7]{BuckinghamM22}, which shows that the maximal domain under consideration is exactly that mapped by $Y=y^{1/3}$ onto the interior of the right ``wing'' of the ``bow-tie'' region $\mathcal{B}$ in the $Y$-plane (see Figure~\ref{fig:DensityPlots}, left-hand panel). The topological structure of $K'$ remains the same under continuation, and hence the claimed structure of $K'$ holds throughout this domain in case (i) as that is the case for $0<y<y_\mathrm{c}$. The boundary of the domain consists of the imaginary segment between $y=\pm 2\mathrm{i}/
\sqrt{27}$ and a certain Schwarz-symmetric curve in the right half-plane connecting those two imaginary endpoints via the positive real value $y_\mathrm{c}\approx 0.29177$; see the right-hand panel of Figure~\ref{fig:DensityPlots}.\looseness=-1

Given the proven structure of $K'$ in case~(i), it is clear that $K'$ divides the complex $\eta$-plane into three disjoint regions, consistent with the \emph{basic structure theorem}~\cite[p.~37]{Jenkins58}. As each of these regions has exactly one pole of $f(\eta,y,c)\,\mathrm{d}\eta^2$ on its boundary (including the point at infinity in a suitable local coordinate), they are all \emph{end domains}, which means that they are conformally mapped by the primitive $\eta\mapsto \int^\eta h_\eta(\eta',y,c)\,\mathrm{d}\eta'$ onto a half-plane with a~vertical boundary. Since $\operatorname{Re}(h(\eta,y,c))=0$ on the boundary of each end domain because it is a subset of $K'\subset K$, the vertical boundary of the image is exactly the imaginary axis, and hence there can be no other points with $\operatorname{Re}(h(\eta,y,c))=0$ in the interior of each end domain. This proves that $K=K'$, which establishes the key properties of the level curve $K$ asserted in the statement of the lemma.

Introducing a contour arc $\Sigma_{0,2}$ joining $\eta=0$ with $\eta=\mathrm{i} s_2$ and an unbounded arc $\Sigma_{1,\infty}$ with finite endpoint $\eta=\mathrm{i} s_1$ such that $\Sigma_h:=\Sigma_{2,3}\cup\Sigma_{0,2}\cup\Sigma_{0,1}\cup\Sigma_{1,\infty}$ is a simple contour, a function $\eta\mapsto h(\eta,y,c_1(y))$ is well-defined up to an integration constant possibly depending on $y$ by contour integration of $h_\eta(\eta,y,c_1(y))$ in the simply connected domain~${\mathbb{C}\setminus\Sigma_h}$. Since $h_\eta(\eta,y,c)$ is integrable at all three of the roots $\eta=\mathrm{i} s_j$, we can and will choose the integration constant so that $\operatorname{Re}(h(\mathrm{i} s_3,y,c_1(y)))=0$. It follows from~\eqref{eq:Boutroux} and the reality of the residue of $h_\eta(\eta,y,c_1(y))$ at $\eta=\infty$ that $\operatorname{Re}(h(\eta,y,c_1(y)))$ is a function harmonic on the larger domain $\mathbb{C}\setminus(\Sigma_{2,3}\cup\Sigma_{0,1})$. Recalling that $\Sigma_{2,3}$ and $\Sigma_{0,1}$ have been chosen to agree~with arcs of $K'=K$, the function $\eta\mapsto\operatorname{Re}(h(\eta,y,c_1(y)))$ is continuous except at ${\eta=0}$.
\end{proof}

\subsection*{Acknowledgements}

R.J.~Buckingham was supported by the National Science Foundation under Grant
DMS-2108019. P.D.~Miller was supported by the National Science Foundation under
Grants DMS-1812625 and DMS-2204896.

\pdfbookmark[1]{References}{ref}
\LastPageEnding

\end{document}